\documentclass[prl,superscriptaddress,twocolumn,aps,showpacs,amsmath,amssymb,floatfix]{revtex4-2}

\usepackage{xr}
\usepackage{amsmath}
\usepackage{amsfonts, amssymb,amsxtra}
\usepackage[]{graphicx}
\usepackage{grffile}
\usepackage{amsfonts}
\usepackage{framed}
\usepackage{bbm}
\usepackage{braket}
\usepackage{xcolor}
\usepackage{chngcntr}
\usepackage{comment}
\usepackage[normalem]{ulem}

\counterwithout{equation}{section} 

\usepackage{bm}
\usepackage{epsfig}
\usepackage{blindtext}
\usepackage[most]{tcolorbox}
\usepackage{floatrow}
\usepackage{verbatim}
\usepackage{color}
\usepackage{diagbox}
\usepackage{outlines}
\usepackage{cancel}
\usepackage{simpler-wick}

\newcommand{\Imp}{\operatorname{Im}}
\newcommand{\Rep}{\operatorname{Re}}

\usepackage[colorlinks=true]{hyperref}

\def\XXint#1#2#3{{\setbox0=\hbox{$#1{#2#3}{\int}$}
     \vcenter{\hbox{$#2#3$}}\kern-.5\wd0}}

\newcommand{\blue}[1]{\textcolor{blue}{#1}}

\newcommand{\avg}[1]{\langle #1 \rangle}

\begin{document}
\title{Interplay of charge and spin fluctuations in a Hund's coupled impurity}
\author{Victor Drouin-Touchette}
\affiliation{Center for Materials Theory, Department of Physics and Astronomy, Rutgers University, Piscataway, NJ 08854 USA}
\author{Elio J. K\"onig}
\email{elio.j.koenig@gmail.com}
\affiliation{Max-Planck Institute for Solid State Research, 70569 Stuttgart, Germany}
\author{Yashar Komijani}
\affiliation{Department of Physics, University of Cincinnati, Cincinnati, Ohio 45221-0011, USA}
\author{Piers Coleman}
\affiliation{Center for Materials Theory, Department of Physics and Astronomy, Rutgers University, Piscataway, NJ 08854 USA}
\affiliation{Department of Physics, Royal Holloway, University of London, Egham, Surrey TW20 0EX, UK}
\date{\today}

\begin{abstract}
In Hund’s metals, the local ferromagnetic interaction between orbitals leads to an emergence of complex electronic states with large and slowly fluctuating magnetic moments. Introducing the Hund’s coupled mixed valence quantum impurity, we gain analytic insight into recent numerical renormalization group studies. We show that valence fluctuations drastically impede the development of a large fluctuating moment over a wide range of temperatures and energy, characterized by quenched orbital degrees of freedom and a singular logarithmic behavior of the spin susceptibility $\chi_{\rm sp}''(\omega) \propto [\omega \ln(\omega/T_K^{\rm eff})^2]^{-1}$, closely resembling power-law scaling $\chi_{\rm sp}''(\omega) \sim \omega^{-\gamma}$. Finally, we outline how such singular spin fluctuations can play an important role in generating a superconducting state through Hund's driven Cooper pairing.
\end{abstract}

\maketitle


\emph{\blue{Introduction -}} The concept of Hund's metals was first introduced in the context of iron-based superconductors \cite{haule2009coherence, yin2012fractional, fanfarillo2015electronic}, with the nomenclature now being extended to include the ruthenates \cite{dang2015electronic, deng2016transport, wang2020hund}. In both cases, the local physics is characterized by electronic shells that are one filling away from half-filling \cite{de2011hund, georges2013strong}. Although the onsite Coulomb interaction $U$ is the largest scale, its effect is overshadowed by that of the ferromagnetic inter-orbital Hund interaction \cite{stadler2019hundness}. This class of materials presents  an intermediate paramagnetic regime dominated by slowly fluctuating high-spin configurations \cite{werner2008spin, watzenbock2020characteristic, hansmann2010dichotomy} and largely suppressed Fermi liquid coherence scales, leading to anomalous transport properties \cite{hardy2013evidence, hardy2016strong, yang2017observation}. It has been speculated that Cooper pairing emerges out of this intermediate state in iron-based compounds \cite{stewart, hosono2018recent, puetter2012identifying, hoshino2015superconductivity, vafek2017hund, cheung2019superconductivity, coleman2020triplet}.  

The link between large moments generated by Hund's coupling and the exponential reduction of Fermi liquid scales in Kondo impurity models has been studied extensively \cite{schrieffer1967kondo, okada1973singlet,nozieres1980kondo, jayaprakash1981two,nevidomskyy2009kondo, drouin2021emergent}. However, the physical valence of Fe or Ru atoms in Hund's metals deviates from half-filled shells. This has led to a vigorous interest in doped multiorbital models, where an intermediate coupling non-Fermi-liquid fixed point was pointed out through analytical \cite{aron2015analytic} and NRG studies \cite{horvat2019non,wang2020global, walter2020uncovering} for the $S=1$, three orbital system. Most of these studies were performed for exactly two electrons among three orbitals; few explicitly addressed charge fluctuations out of this state \cite{stadler2019hundness,ryee2022frozen}. 

Motivated by the potentially new physics in the mixed valence regime, here we extend our previous work on the Hund-Kondo impurity \cite{drouin2021emergent}. This, coupled with the unique property of the large-N self-consistent equations which enable a direct access to real frequency correlation functions \cite{komijani2018model}, has led us to study in detail the dynamical properties of the intermediate regime generated by Hund's coupling.   




\begin{figure}[t]
\includegraphics[width=\textwidth]{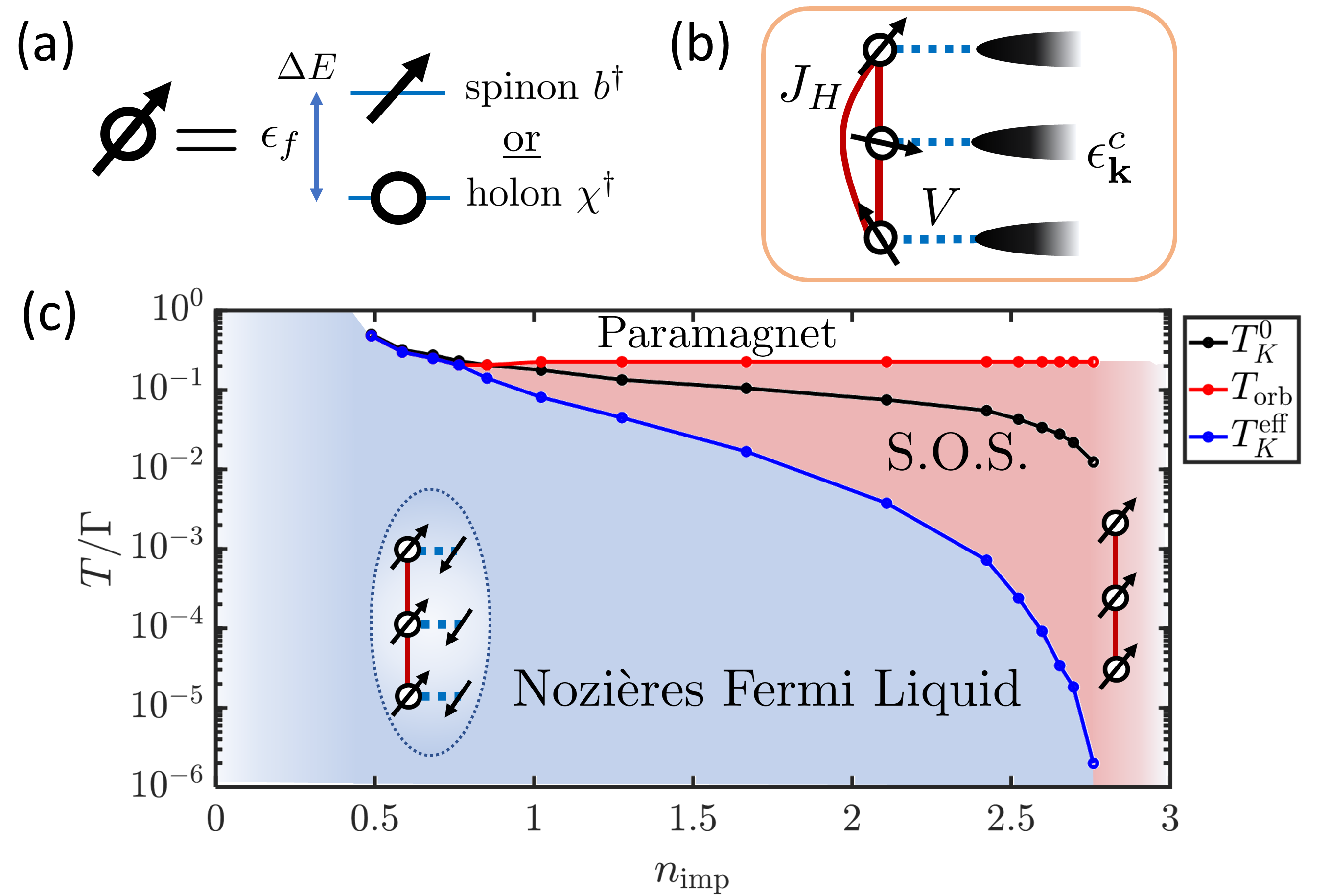} 
\caption{{\bf (a-b)} Schematic mixed valence states and of the Hund-coupled three-orbital Anderson model, referring to Eq.~\eqref{eq-model}. {\bf (c)} Phase diagram obtained as a function of total impurity filling $n_{\rm imp}$ (left side is holon dominated). $T_{\rm orb}$ and $T_K^{\rm eff}$ are crossover temperatures where the specific heat $c_v$ has a local maxima, as in Fig.~\ref{fig:moment}. Generically, screening occurs in two steps. The formation of a large emergent fluctuating moment, while orbital degrees of freedom are quenched, occurs at $T_{\rm orb}$. In this spin-orbital separated (S.O.S.) regime, we see a Curie-like spin susceptibility $\chi_{\rm sp} \sim \mu^2/T$ while the orbital susceptibility reaches a plateau. Then, at $T_K^{\rm eff} \ll T_K^0$, with $T_K^0$ the bare Kondo temperature for $J_H = 0$, the large moment is screened and forms a local Fermi liquid. $\Gamma = \pi \rho V^2$ is the bare hybridization width.}
\label{fig:phasediagram}
\end{figure}


In this Letter, we show that the treatment of charge and spin dynamics on equal footing within the dynamical large-N approach \cite{rech2006schwinger, komijani2018model, komijani2019emergent, wang2020quantum, shen2020strange, anyon2020komijani, wang2021nonlocal, drouin2021emergent, han2021schwinger, ge2022emergent} leads to new insight into the dynamical properties of hole doped multi-orbital impurities. Based on computed thermodynamic quantities, we unveil the complete phase diagram [see Fig.~\ref{fig:phasediagram}] as a function of impurity occupancy. Furthermore, we characterize the large emergent moment regime \cite{drouin2021emergent} as one with spin-orbital separation \cite{stadler2015dynamical}. The spin susceptibility shows logarithmic corrections due to the extremely slow approach to Kondo screening, reminiscent of the low-energy properties of the underscreened Kondo model \cite{varma2002singular, coleman2003singular,mehta2005regular, coleman2005quantum}. At a strongly renormalized Kondo temperature $T_K^{\rm eff} \ll T_K^0$, the charge degrees of freedom eventually gap out, resulting in a local Fermi liquid. The scaling of the local spin susceptibility in the spin-orbital separated regime is found to exhibit quasi-power-law scaling due to the unusually strong logarithmic corrections.


\emph{\blue{Model -}} We consider a degenerate three-orbital impurity model, in analogy to the $t_{2g}$ orbital subset generated due to the tetrahedral environment around the iron atoms in Fe-based Hund's metals. Mixed-valence states are included via a Hund-Anderson model. We take the limit of $U \rightarrow \infty$, such that the Coulomb interaction enforces a ``no double occupancy" rule for each of the $m$ orbitals. This is a valid limit away from exactly half-filling where we can focus on the role of Hund's coupling. This is enforced through the use of Hubbard operators \cite{coleman2015introduction} for each orbital, which transform the $K$ empty states $\ket{d^0: a,m}$ into the $N$ magnetic states $\ket{d^1: \alpha,m}$ filled with local $d$ electrons for each orbital $m$. In the large-N formalism, local moments transform as spin-$S$ representations of SU(N), and there are $K = 2S$ electronic channels present to maintain perfect screening. Schwinger bosons $X_{\alpha, \beta}^{(m)}  = b_{m\alpha}^{\dagger} b_{m\beta}$ are used to express the spin degrees of freedom through spinons $b^{\dagger}$ which form a symmetric representation of the spins. Together with the use of slave fermions (holons $\chi^\dagger$), Hubbard operators can faithfully be represented as

\begin{equation}
\begin{split}
X_{\alpha, a}^{(m)} &\equiv \ket{d^1: \alpha, m} \bra{d^0: a, m} = b_{m\alpha}^{\dagger} \chi_{ma} \,, \\
X_{a,\alpha}^{(m)} &\equiv \ket{d^0: a, m}\bra{d^1: \alpha, m} =  \chi_{ma}^{\dagger} b_{m\alpha} \,,\\
X_{a,b}^{(m)} &\equiv \ket{d^0: a, m} \bra{d^0: b, m}= \chi^{\dagger}_{ma} \chi_{mb} \,, \\
X_{\alpha, \beta}^{(m)} &\equiv \ket{d^1: \alpha, m} \bra{d^1: \beta, m} = b_{m\alpha}^{\dagger} b_{m\beta} \,,
\end{split} \label{eq-hubbard}
\end{equation}

\noindent while the Hamiltonian is itself expressed as $H = \sum_m H_c^{(m)} + \sum_m H_{\rm K}^{(m)} + H_{\rm H}$, with individual terms 

\begin{subequations}
\begin{align}
H_c^{(m)} &=  \sum_{\bf k} \epsilon_{\bf k}^c c^\dagger_{{\bf k} m \alpha a }   c_{{\bf k} m \alpha a} \;, \\
H_{\rm K}^{(m)} &=  V( c^\dagger_{0 m \alpha a} X_{a, \alpha}^{(m)} + X_{\alpha, a}^{(m)} c_{0 m \alpha a}) + \epsilon_f X_{a , a}^{(m)}  \;, \\
H_{\rm H} &= - \frac{J_H}{N} \sum_m  X_{\alpha \beta}^{(m)} X_{\beta \alpha }^{(m+1)} \;.
\end{align} \label{eq-model}
\end{subequations}

A visual representation of the mixed valence states and the Hund-Anderson model is presented at Fig.~\ref{fig:phasediagram} (a). Here, $V$ is the hybridization between conduction electrons and the Hubbard operators corresponding to adding or removing an impurity electron ($\Gamma = \pi \rho V^2$ is the bare hybridization width). We denote the energy of a hole as $\epsilon_f$, which, when tuned, leads to different $n_{\rm imp}$ as the average valence of each state is changed. A more realistic model which includes crystal field splitting effect between the orbitals could be implemented by tuning $\epsilon_f^{(a)}$ for each orbital independently \cite{werner2009metal,lanata2013orbital}. This would lead to orbital differentiation \cite{kugler2019orbital, kugler2021orbital}, and is beyond the scope of our work \footnote{One could include realistic crystal field splitting effects in our model by breaking the degeneracy of the three orbitals. The isotropic holon energy term in Eq.~\eqref{eq-model} would become $\epsilon_f [X^{(1)}_{a,a} + X^{(1)}_{a,a}] + (\epsilon_f + \Delta_{cf})X^{(3)}_{a,a}$, which leads to an additional self-energy equation for the holons. We expect that a finite $\Delta_{cf}$ will lead to orbital differentiation, with different $T_{\rm orb}$ crossover temperatures for each orbital subset, but that the S.O.S. regime would remain. The detailed effects of such perturbations are beyond the scope of our work.}. The Hund's term $H_{\rm H}$ can be treated through a Hubbard-Stratonovich decoupling in the hopping channel:

\begin{equation}
    H_{\rm H} \rightarrow \sum_m [\bar{\Delta}_m (b^{\dagger}_{m+1,\alpha} b_{m \alpha}) + h.c.] + \frac{N |\Delta_m|^2}{J_H} \,. \label{eq-stratonovich}
\end{equation}

A mean-field equation relates $J_H$ to the spinon gap $\Delta$, such that generically, $\Delta/J_H = \avg{\sum_{m} b^{\dagger}_{m+1} b_m}$ \cite{drouin2021emergent}. If $J_H$ is large enough to generate a finite $\Delta$, their relationship will be such that $N\Delta = 2S n_{\rm imp} J_H$ for $T \gg T_K^0$. Furthermore, the total charge 

\begin{equation}
Q_m = \sum_{\alpha} b^{\dagger}_{m\alpha} b_{m\alpha} + \sum_a \chi^{\dagger}_{m a} \chi_{m a} = 2S \,,\label{eq-conscharg}
\end{equation}

\noindent at each orbital commutes with the Hubbard operators and is a conserved quantity, setting the size of the local moments. This is a constraint on the spinons/holons, enforced through a common Lagrange multiplier $\lambda$. In the large-N limit, the dynamics of holons and spinons is dominated by the non-crossing Feynman diagrams, which leads to the self-energy equations

\begin{equation}
\Sigma_{\chi} (\tau) = g_c(-\tau) G_B (\tau) \; , \; \Sigma_{B} (\tau) = - k g_c(\tau) G_{\chi} (\tau) \;,\label{eq-selfcons}
\end{equation}

\noindent where $k = K/N = 2S/N$ and $g_{c,0} (z) = \sum_{\bf k} [z - \epsilon^c_{\bf k}]^{-1}$ corresponds to the conduction electron's bare propagator for imaginary frequencies $z$. Eqs.~\eqref{eq-selfcons} are solved self-consistently together with the Dyson equations $G_b (z) = \sum_m G_B(m,z) = \sum_p [z - \epsilon_p - V^2 \Sigma_b (z)]^{-1}$ and $G_{\chi} (z) = [ z - \epsilon_f^{\ast} - V^2 \Sigma_{\chi} (z)]^{-1}$. Here, we used the following definitions: $G_B(m,z)$ is the spinon's Green's function on orbital $m$, $\epsilon_p = (\lambda -2\Delta \cos{p})$ is the energy of the spinon states and $\epsilon_f^{\ast} = \lambda + \epsilon_f$ is the holon energy. We also defined the chirality $p = 0, \pm 2\pi/3$ which denotes the chosen Hund energy levels such that $p=0$ is the aligned state (maximum total spin). $\lambda$ and $\Delta$ are adjusted to fit the constraint of Eq.~\eqref{eq-conscharg} and the mean-field relation between $J_H$ and $\Delta$. 

Thermodynamic and dynamical observables are obtained from the Green's functions \cite{rech2006schwinger, komijani2018model, drouin2021emergent}. Notably, the impurity entropy $S_{\rm imp}(T)$ can be extracted exactly in the large-N limit \cite{coleman2005sum, lebanon2006conserving} and is given by 

\begin{equation}
\begin{split}
    &S_{\rm imp} = - {\rm Tr} \int \frac{d\omega}{\pi} \Big( \frac{\partial n_{B}}{\partial T} [{\rm Im} \, \ln (-G_{B}^{-1}) + G'_{B} \Sigma_{B}''] \\
    &\; \; + \gamma \frac{\partial n_{F}}{\partial T} [{\rm Im} \, \ln (-G_{\chi}^{-1}) + G'_{\chi} \Sigma_{\chi}'' - g''_{c,0} \tilde\Sigma_{c}'] \Big) \,, \label{eq-entropy}
\end{split}
\end{equation}

\noindent where the trace is over all $\alpha$ and $a$ spin and channel indices and chiralities $p$, and $\tilde{\Sigma}_c (\tau)= G_{\chi}(-\tau) G_B (\tau)$.  $G'_{\xi}$ is the real part of the retarded Green's function while $\Sigma_{\xi}''$ is the imaginary part of the self-energy for the corresponding fields. From this closed form, we can extract the specific heat as $c_{v, \rm imp} = T \partial S_{\rm imp} /\partial T$.

\begin{figure}[t]
\includegraphics[width=\textwidth]{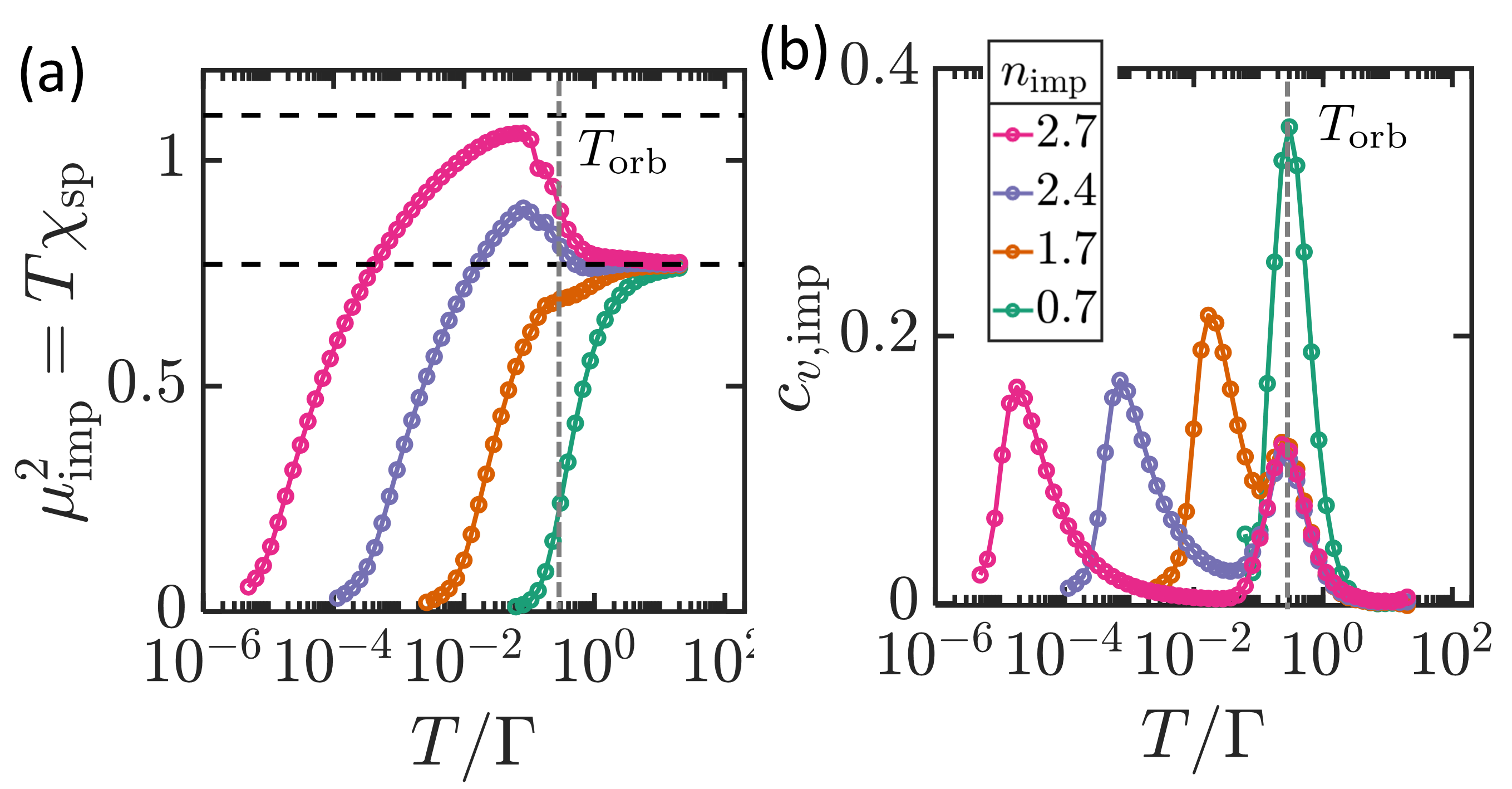} 
\caption{\textbf{(a)} The impurity's local moment $\mu_{\rm imp}^2$ obtained from the spin susceptibility, for varied impurity occupations $n_{\rm imp}$. The large emergent moment, seen as a low-temperature higher plateau, is destroyed as more holons are added, following the trend of Fig.~\ref{fig:phasediagram}. \textbf{(b)} The specific heat $c_{v, \rm imp}$, obtained from the closed form of the entropy for the multiorbital Anderson model, see Eq.~\eqref{eq-entropy}. Both $T_{\rm orb}$ and $T_{K}^{\rm eff}$ are crossovers associated with $c_v$ peaks. Dashed lines are high and intermediate temperature limits for the $\mu_{\rm imp}^2$, obtained in Ref.~\cite{drouin2021emergent}.}
\label{fig:moment}
\end{figure}


\emph{\blue{Summary of results -}} In the absence of Hund's coupling, one simply recovers three copies of infinite-U Anderson models. In this case, the presence of holons (decreasing $n_{\rm imp}$) increases the bare Kondo temperature $T_K^0  \sim D e^{-1/J_K^0 \rho} \sim D \exp{(-|\epsilon_f|/\Gamma)}$ for a fixed hybridization width $\Gamma$, conduction electron bandwidth $D$ and effective holon energy $\epsilon_f^{\ast}$. The holon occupation number $n_{\chi}$ can be absorbed through a Gutzwiller renormalization of the hybridization $V \rightarrow \tilde{V} \sim V \sqrt{\avg{n_{\chi}}}$ \cite{coleman2015introduction}, such that as $n_{\chi} \rightarrow 0$, $T_K^0 \rightarrow 0$ exponentially. Solving the self-energy equations for $J_H=0$ leads to this expected trend, shown in black in Fig.~\ref{fig:phasediagram} (c) \cite{suppmaterials}. 

For a finite Hund's coupling, the situation changes drastically. The emergence of an intermediate large moment phase in the Kondo limit is consistent with our previous work on the Hund-Kondo model \cite{drouin2021emergent}. We can now connect this phase continuously throughout the hole-doped regime.  At some critical holon doping, there is no longer enough local moments to lock together - the two-step Kondo screening reverts to a single step Kondo crossover.  The obtained phase diagram of Fig.~\ref{fig:phasediagram} is consistent with other NRG+DMFT studies \cite{stadler2019hundness} of hole doped multiorbital impurity models.

In Fig.~\ref{fig:moment}, we present the thermodynamic measurements of the total impurity's magnetic moment $\mu^2 \sim T\chi_{\rm sp}$ as well as the impurity specific heat for select impurity occupations $n_{\rm imp}$ throughout the entire temperature range. For Kondo-like systems ($n_{\rm imp} = 2.7$) there is a clear non-monoticity in the local moment, signaling the intermediate formation of an emergent large moment due to Hund's coupling. As the holon occupancy increases, there are less spinons in the system and the emergent moment can no longer form - the shoulder disappears at $n_{\rm imp} \approx 0.7$. In the specific heat, this disappearance of the intermediate phase is seen as the low and high temperature crossover features merge to become one as $n_{\rm imp} = 0.7$, where single-step Kondo screening occurs. 

Previous works \cite{stadler2015dynamical,stadler2019hundness, stadler2019model, walter2020uncovering, wang2020global} have characterized the intermediate large moment phase as a regime with spin-orbital separation. We can take advantage of the Hubbard operators' description in terms of spinons and holons to write composite orbital operators. This procedure would not be possible with virtual holons, as obtained in past treatments of the Hund-Kondo model \cite{drouin2021emergent}. The finite holon occupancy in the Hund-Anderson model leads to a well-defined orbital degrees of freedom. Starting with the total impurity spin operators at imaginary time $\tau$, described as $S_{\alpha \beta} (\tau) = \sum_m X_{\alpha \beta}^{(m)} (\tau) = \sum_m b^{\dagger}_{m \alpha} (\tau) b_{m \beta} (\tau)$ ($\alpha, \beta$ are SU(N) spin states), we then harness the SO(3) orbital symmetry and describe impurity orbital operators $\hat{L}_{\gamma}$ ($\gamma = x,y,z$ corresponding to the three degenerate orbitals) as $\hat{L}_{\gamma} = (1/NK) \sum_{mm' \alpha a} X^{(m)}_{\alpha a} (L_{\gamma})_{ m m'} X^{(m')}_{a \alpha}$. The $L_{\gamma}$ are generators of the SO(3) group \footnote{Another definition of $\hat{L}_{\gamma} = \sum L_{\gamma}^{m m'} X_{ab}^{m} X_{ba}^{m'}$ purely in terms of orbital charge can generically be written. For the antisymmetric $L_{\gamma}$ matrices, this leads to an identically zero operator.} such that $(L_{\gamma})_{mm'} = i \epsilon_{\gamma m m'}$ with $\epsilon_{ijk}$ the anti-symmetric Levi-Civita tensor. The spin and orbital susceptibilities in the large-N limit can thus be expressed as 

\begin{figure*}[t]
\includegraphics[width=\textwidth]{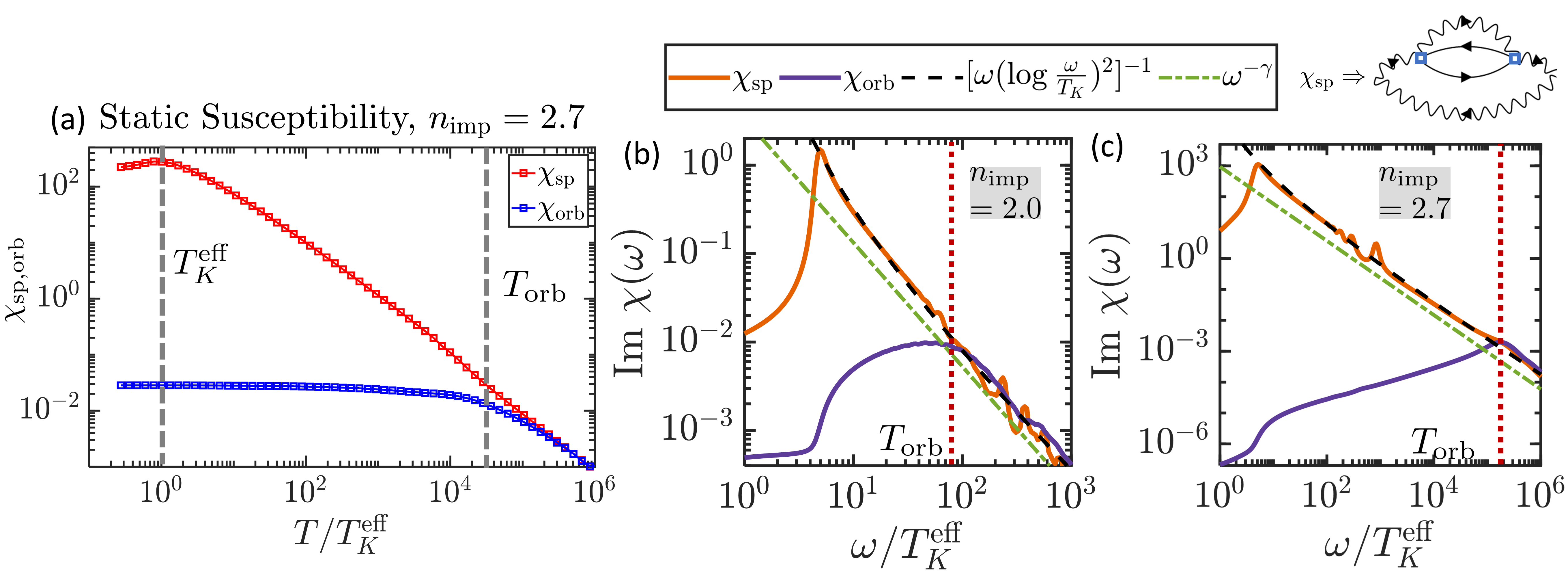} 
\caption{{\bf (a)} Static spin and orbital susceptibilities for $n_{\rm imp} = 2.7$, showing a clear separation at $T_{\rm orb}$ below which the Curie-like $\chi_{\rm sp} \sim \mu^2/T$ spin susceptibility is in stark constrast to the constant $\chi_{\rm orb}$. {\bf (b-c)} Imaginary part of the dynamical spin and orbital susceptibilities, Eqs.~\eqref{eq-chiall0}, presented for different occupations of the three-orbital impurity: $n_{\rm imp} = 2$  and $n_{\rm imp} = 2.7$ from left to right. The value of the Hund coupling $J_H/D = 0.37$ is fixed, and results are presented for $T \simeq 0.1 T_K^{\rm eff}$. The dashed black line corresponds to the derived second-order perturbation scaling of the spinon susceptibility, Eqs.~\eqref{eq-scale}, while the dot-dashed green line is the quasi-power-law form of Eq.~\eqref{eq-gammamain} (with small offset for readability). At the top right, we show the second order contribution to $\chi_{\rm sp}$ which leads to the logarithmic scaling. Conduction electron (spinon) Green's functions are represented as solid (wavy) lines.}
\label{fig:susceptibility}
\end{figure*}

\begin{subequations}
\begin{align}
\chi_{\rm sp} (\tau) &= \frac{1}{N^2} \sum_{\gamma} \sum_{\alpha \beta} \langle S_{\alpha \beta}^{(\gamma)} (\tau) S_{\beta \alpha}^{(0)} (0) \rangle_c \notag \\
& \rightarrow \sum_m G_B (m, \tau) G_B (- m, - \tau) \,, \label{eq-chisp0} \\
\chi_{\rm orb} (\tau) &= \frac{1}{3} \sum_{\gamma}  \langle \hat{L}_{\gamma} (\tau) \hat{L}_{\gamma} (0) \rangle_c \notag \\
& \rightarrow \frac{1}{K} \Big( \frac{\Delta}{J_H} \Big)^2 G_{\chi} (\tau) G_{\chi} (- \tau) \,, \label{eq-chiorb0}
\end{align} \label{eq-chiall0}
\end{subequations}

\noindent where $\langle \cdots \rangle_c$ denotes averages over connected diagrams. The derivation of the expression for the spin susceptibility is identical to our previous work \cite{drouin2021emergent}. The expression for the orbital susceptibility is obtained through two essential steps. Firstly, after the $\hat{L}_{\gamma}$ operators are represented in terms of Hubbard operators, Wick contractions over the bosonic and fermionic degrees of freedom leaves only two relevant contributions in the large-N limit. Secondly, the absence of holon interorbital hopping leads to many terms being zero. Summing over $\gamma = x,y,z$ leads to the quoted result. The full expression for the dynamical susceptibilities in real frequency are presented in the supplementary materials \cite{suppmaterials}, as well as further details on this derivation. Note that the orbital susceptibility has a $1/K$ factor reduction compared to $\chi_{\rm sp}$; we nevertheless can plot $K \chi_{\rm orb}$ and obtain valuable insight. The static components of these susceptibilities, $\chi_{\rm orb} (\tau = 0)$ and $\chi_{\rm sp} (\tau = 0)$, are shown in panel (a) of Fig.~\ref{fig:susceptibility}. The clear splitting of both susceptibilities at $T_{\rm orb}$, and the subsequent plateau in $\chi_{\rm orb}$, signals the formation of the large moment and the separation of spin and orbital scales (S.O.S). Throughout this regime, orbital and charge fluctuations are nearly frozen while the spin susceptibility remains Curie-like.


\emph{\blue{Dynamical susceptibilities -}} The emergent moment regime has clear thermodynamic attributes, as described above (see Fig.~\ref{fig:moment}). Further insight into this phase is provided by the dynamical spin and orbital susceptibilities, as defined in Eqs.~\eqref{eq-chiall0}. We show these in Fig.~\ref{fig:susceptibility} for two different total impurity valences. It can be clearly seen that at high frequencies, both spin and orbital degrees of freedom fluctuate freely. For $\omega < T_{\rm orb}$, the lower-energy high-spin configurations split off, which is associated with the separation of the spin and orbital dynamical susceptibility. In this regime, the charge fluctuations freeze out and the valence stabilizes below $T_{\rm orb}$; this quenching of orbital degrees of freedom leads to the decrease in $\chi_{\rm orb}''$ with respect to $\chi_{\rm sp}''$.

From Fig.~\ref{fig:susceptibility}, we see that, for many decades in frequency between $T_K^{\rm eff}$ and $T_{\rm orb}$, the spin susceptibility seems to grow in a power-law $\chi_{\rm sp} \sim \omega^{-\gamma}$ (dot-dashed green line). In Kondo impurity problems, such behavior is often indicative of non-Fermi-liquid fixed points \cite{walter2020uncovering, wang2020global}, for example in the 2-channel spin-$1/2$ Kondo model \cite{relevance1992affleck, exact1993affleck, coleman1995simple, georges1995solution,parcollet1997transition, parcollet1998overscreened}. Closer examination reveals that this is not the case in this system, having maintained perfect screening ($k=q$) throughout. Instead, we find a good agreement at intermediate temperatures and frequencies with the scaling

\begin{align}
\chi_{\rm sp}''(\omega) &= \frac{(J_K^{\rm eff} \rho )^2}{\omega} \propto \Big( \omega \Big[\ln \Big( \frac{\omega}{T_K^{\rm eff}} \Big) \Big]^2  \Big)^{-1} \;. \label{eq-scale}
\end{align}

Here we cover the basic steps of this derivation and leave the details for the supplementary materials \cite{suppmaterials}. Firstly, we solve a single iteration \cite{drouin2021emergent} of the self-energy equations of Eq.~\eqref{eq-selfcons} analytically, starting from the bare Green's functions $G_{\xi,0} (z)$. This leads to an expression for the renormalized Kondo temperature $T_K^{\rm eff}$. Furthermore, for $T_K^{\rm eff} \leq \max(\omega, T) \leq T_{\rm orb}$, we can map the mixed valence problem onto a Kondo problem, leading to an effective holon propagator $\tilde{G}_{\chi} (\omega) = - J_K^{\rm eff}(\omega)$, with 

\begin{equation}
    \frac{1}{\rho J_K^{\rm eff} (\omega)} \simeq \ln \Big( \frac{\max (\omega, T)}{T_K^{\rm eff}} \Big) \;. \label{eq-jkeff0}
\end{equation}

Secondly, after having obtained this effective running Kondo coupling, we proceed in a second-order perturbation in $J_K^{\rm eff}$ of the spinon bubble of the spin susceptibility \cite{koller2005singular}. This is shown in the top right of Fig.~\ref{fig:susceptibility}. Blue boxes corresponds to factors of $\rho J_K^{\rm eff}$, and the calculation of this diagram leads to the scaling presented in Eq.~\eqref{eq-scale}. One can see in Fig~\ref{fig:susceptibility} that it agrees perfectly within the intermediate regime $T_K^{\rm eff} < \omega < T_{\rm orb}$ with only $T_K^{\rm eff}$ as an input parameter. A downturn is observed at lower frequencies consistent with $\Imp \chi_{\rm sp} \propto \omega$ in the Fermi liquid regime. This scaling holds for all $n_{\rm imp}$ of the phase diagram where the SOS phase is present. 

In the SOS regime, the large separation of scales between $T_K^{\rm eff}$ and $T_{\rm orb}$ leads to a peculiar observation about Eq.~\eqref{eq-scale}. For intermediate frequencies, we find that a quasi-power-law form for the spin susceptibility,

\begin{equation}
    \chi_{\rm pwl} \sim \omega^{-\gamma} \, \qquad \text{and} \qquad \gamma = 1 - 2/\ln\Big(\frac{T_K^{\rm eff}}{\mathcal{D}}\Big) \,,\label{eq-gammamain}
\end{equation}

\noindent with $\mathcal{D} = \min (\Gamma, T_{\rm orb})$, is indistinguishable from the form with the logarithmic correction. These two start to deviate as one gets to very small frequencies $\omega \ll \mathcal{D}$ \cite{suppmaterials}, which results in the upturn seen close to $T_K^{\rm eff}$ in Fig.~\ref{fig:susceptibility}. For very small $T_K^{\rm eff}/T_{\rm orb}$, due to strong Hund's coupling and the resulting nearly frozen charge fluctuations, the slow logarithmic scaling presents itself as this quasi-power-law for many decades in frequency. We find that, for a given fixed $J_H/D = 0.37$, $\gamma \simeq 1.2$ for $n_{\rm imp} = 2.7$ and $\gamma \simeq 1.4$ for $n_{\rm imp} = 2.0$. This exponent $\gamma$ changes continuously as $n_{\rm imp}$ is varied.

We note that $\chi_{\rm sp}'' \sim \omega^{-1.2}$ was seen in a different but related model \cite{wang2020global, walter2020uncovering}, and was invoked in phenomenological modeling of the spin-fluctuation-induced Cooper pairing in the iron-based superconductors \cite{lee2018pairing, wu2019pairing}. In those references, the presence of a putative soft boson with $\chi_{\rm sp}'' \sim \omega^{-1.2}$, when included in a Eliashberg approach, led to a superconducting instability with universal properties relevant for the iron-based superconductors, but the origin of this mode was an open question. Our results provide a tentative identification of this mode in terms of Hund's driven Kondo screening.

\begin{figure}[t]
\includegraphics[width=\textwidth]{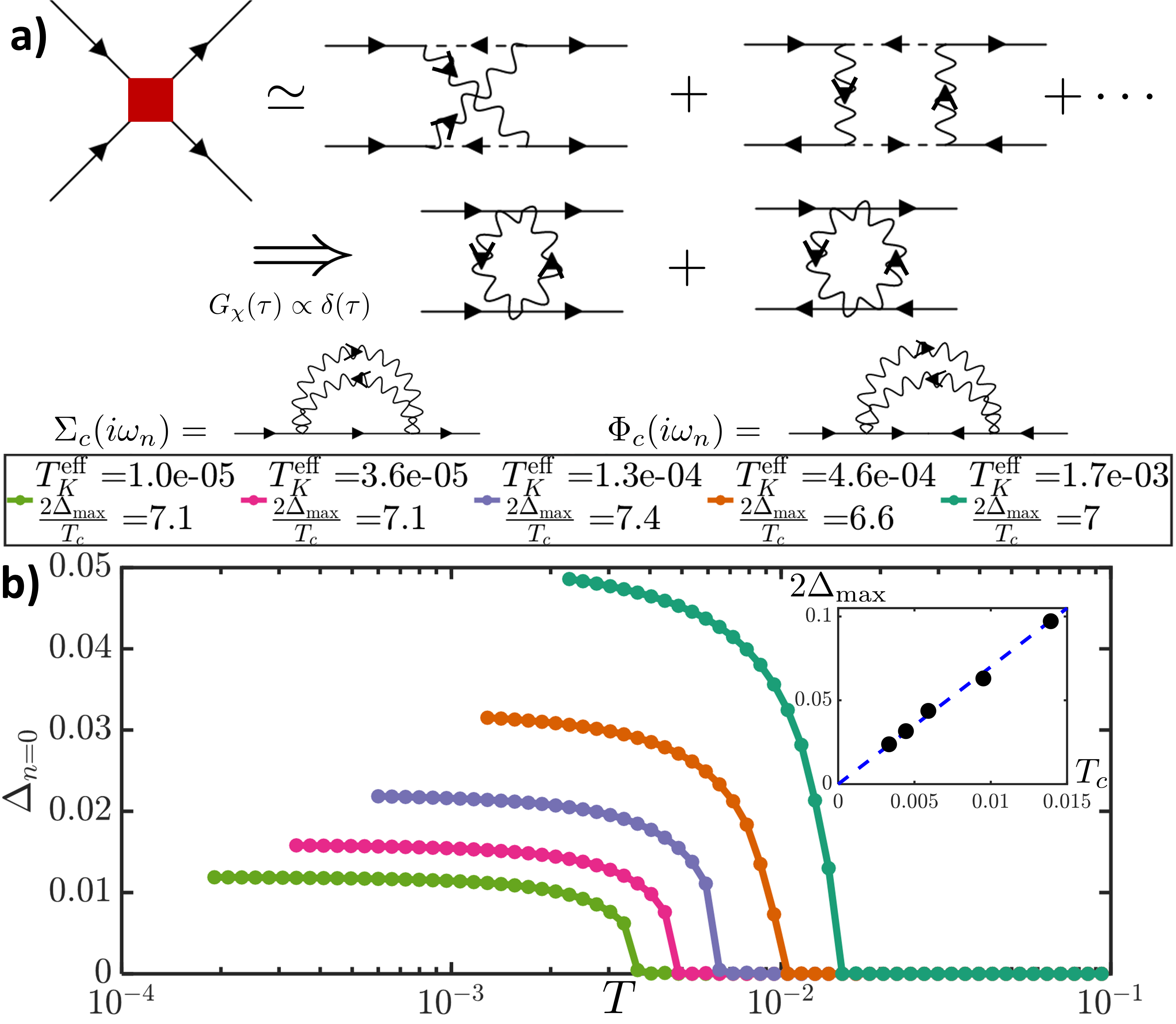} 
\caption{{\bf (a)} In red, interaction kernel for the conduction electrons. In the Schwinger-boson representation, there are two contributions. For temperatures below $T_{\rm orb}$, the holons are nearly instantaneous propagators and the two diagrams can be simplified. A spinon bubble is then the contribution to the pairing vertex $\Phi_c$ and fermionic self-energy $\Sigma_c$. Holon propagators are represented by dashed lines, while conduction electrons (spinons) are represented by full (wavy) lines.. {\bf (b)} Solving Eq.~\eqref{eq-eliashbergSolve} with $\lambda = \chi''_{\rm sp}$, with different $T_K^{\rm eff}$ chosen. The inset shows $2\Delta_{\rm max}$ vs $T_c$ for the presented curves, along with $2\Delta_{\rm max} \propto 7.0 T_c$ in dashed blue.}
\label{fig:eliashberg}
\end{figure}

\emph{\blue{Eliashberg approach -}} We can extend these arguments to the singular local spin susceptibility obtained here. Firstly, we can extract two contributions to the interaction kernel for the conduction electrons at the $O(1/N^2)$ level, shown in Fig.~\ref{fig:eliashberg}: a normal and an anomalous contribution, respectively. In the S.O.S. regime, the quenching of the local moments, which also leads to the apparent quasi-power-law behavior, means that the holon's propagators can be approximated as instantaneous $G_{\chi} \propto \delta(\tau)$ \footnote{In reality, $G_{\chi}$ has a frequency dependent behavior, as we saw in the rest of the paper, but it can be approximated as nearly instantaneous with respect to the other propagators.}. Simplifying the two contributions leads to a pairing vertex and fermionic self-energy of equal magnitude, both occurring through $\chi_{\rm sp}''$ - this acts as our soft boson.


Following the work of Ref.~\cite{lee2018pairing}, we can express the Eliashberg equations \cite{migdal1958interaction, eliashberg1960interactions, combescot1995strong, moon2010quantum, wang2016superconductivity} for the pairing vertex $\Phi_c(\omega_n)$ and the fermionic self-energy $\Sigma_c (\omega_n)$ in a closed form. These can be factorized using the pairing gap function $\Delta(\omega_n) = \Phi(\omega_n) \omega_n/(\omega_n + \Sigma (\omega_n))$, leading to a closed equation for $\Delta (\omega_n) \equiv \Delta_n$:

\begin{align}
    \Delta_n & = \pi T \sum_{\omega_m} \frac{\lambda(\omega_m -  \omega_n)}{\sqrt{\omega_m^2 + \Delta^2_m}} \Big( \Delta_m - \Delta_n \frac{\omega_m}{\omega_n} \Big) \;,  \label{eq-eliashbergSolve}
\end{align}

\noindent where $\omega_n = \pi T(2n+1)$ is the $n$-th fermionic Matsubara frequency and $\lambda(\Omega) \sim \chi''_{\rm sp}(\Omega)$ carries the effect of the spin fluctuation bubble. Note that this form only holds for the intermediate frequency and temperature window of $T_K^{\rm eff} < T,\Omega < T_{\rm orb}$. This can then be used to obtain $\Sigma(\omega_n)$. Solving Eq.~\eqref{eq-eliashbergSolve} shows that as the temperature is lowered, a finite $\Delta_n \neq 0$ develops below $T_c$. In Fig.~\ref{fig:eliashberg} (b), we show the maximum gap, achieved at $n=0$, as a function of temperature. The critical temperature and the maximal gap $\Delta_{\rm max} = \Delta_0 (T \rightarrow 0)$ are seen to scale with $T_K^{\rm eff}$ and $T_c$. For all $T_K^{\rm eff}$ studied, the SOS window is large enough to generate a superconducting state within the Eliashberg approach. Furthermore, we find that $2\Delta_{\rm max}/T_c \sim 7.0 \pm 0.5$ for a wide range of $T_K^{\rm eff}/T_{\rm orb}$, close to the universal value observed in Ref.~\cite{miao2014coexistence}.


\emph{\blue{Conclusion -}} We have shown that the dynamical large-N approach can capture the destruction of the Hund's coupled emergent large moment due to hole doping. Furthermore, we show that the intermediate regime is well described through the concept of spin-orbital separation (SOS) \cite{stadler2015dynamical, stadler2019hundness}. In this phase, the dynamical spin susceptibility has a logarithmic component due to the nearly frozen charge fluctuations, which presents itself as a quasi-power-law for an extended frequency range because $T_K^{\rm eff} \ll T_{\rm orb}$. The non-Fermi-liquid-like features in the emergent moment regime can be continuously connected to the integer valence limit. We have also shown how the singular aspects of this spin susceptibility can be included in a Eliashberg treatment and lead to a superconducting state with quasi-universal properties reminiscent of the iron-based superconductors.

\emph{\blue{Acknowledgments -}} VDT would like to thank Elias Walter, Seung-Sup B. Lee, Andreas Weichselbaum, Jan von Delft, Fabian Kugler, Abhishek Kumar and Tsung-Han Lee for illuminating discussions. We also acknowledge useful discussions with Thomas Sch\"afer. This work was supported by DOE Basic Energy Sciences grant DE-FG02-99ER45790 (VDT, PC) and the Fonds de Recherche Québécois en Nature et Technologie (VDT).


\bibliography{biblio}


\clearpage
\begin{widetext}

\setcounter{equation}{0}
\setcounter{figure}{0}
\setcounter{section}{0}
\setcounter{table}{0}
\setcounter{page}{1}
\makeatletter
\renewcommand{\theequation}{S\arabic{equation}}
\renewcommand{\thesection}{S\arabic{section}}
\renewcommand{\thefigure}{S\arabic{figure}}

\begin{center}
Supplementary materials on \\
\textbf{"Interplay of charge and spin fluctuations in a Hund's coupled impurity"}\\
Victor Drouin-Touchette$^{1}$, Elio J. K\"onig$^{2}$, Yashar Komijani$^{3}$, Piers Coleman$^{1,4}$\\ 
$^{1}$\textit{Center for Materials Theory, Department of Physics and Astronomy, Rutgers University, Piscataway, NJ 08854 USA} \\
$^{2}$\textit{Max-Planck Institute for Solid State Research, 70569 Stuttgart, Germany}\\
$^{3}$\textit{Department of Physics, University of Cincinnati, Cincinnati, Ohio 45221-0011, USA}\\
$^{4}$\textit{Department of Physics, Royal Holloway, University of London, Egham, Surrey TW20 0EX, UK}
\end{center}

These supplements contain the following sections:
\begin{itemize}
    \item[I.] Details on the bare mixed valence problem in the absence of Hund's coupling;
    \item[II.] The derivation of the spin and orbital susceptibilities; 
    \item[III.] Details on the single iteration approach to logarithmic corrections of the Kondo coupling; 
    \item[IV.] The consequence of the logarithmic approach to the Fermi liquid for the dynamical spin susceptibility and the subsequently obtained scaling form; 
    \item[V.] The derivation of the quasi-power-law exponent $\gamma$.
\end{itemize}

\section{I. Infinite-U Anderson Model in the large-N approach}

We here present further details on the black curve from Fig.~1 (c) in the main text, since only once \cite{lebanon2006conserving} has the infinite-U Anderson model been studied in the large-N approach. In this context, the interorbital Hund's coupling $J_H$ is set to $0$. While the self-energy equations are the same, the Dyson equations for the holons and spinons are 

\begin{align}
    G_b (z)^{-1} &= [z - \lambda - V^2 \Sigma_b (z)] \,,\\
    G_{\chi}(z)^{-1} &= [ z - \lambda - \epsilon_f - V^2 \Sigma_{\chi} (z)] \,.
\end{align}

We numerically solve these Dyson equations together with the self-energy self-consistent equations, for a fixed bare hybridization width $\Gamma = \pi \rho V^2$ with $\rho = 1/(2D)$ the conduction electron's density of states, and a fixed electronic bandwidth $D$. 

For such an infinite-U Anderson-model, it is known \cite{lebanon2006conserving} that the characteristic Kondo temperature is given by

\begin{equation}
    T_K^0 = D \Big( \frac{\Gamma}{\pi D} \Big)^{k} \exp \Big\{ -\frac{\pi |\epsilon_f|}{\Gamma} \Big\} \,,\label{eq-tkAnderson}
\end{equation}

\noindent where $k = q = 2S/N$ for this perfectly screened case. We show in Fig.~\eqref{fig:supp1mv} the obtained Kondo temperature as a function of the holon energy level from numerically solving the large-N equations. We can see that, having defined the holon energy in the action as $\epsilon_f \chi_i^{\dagger} \chi_i$, $\epsilon_f > 0$ will result in a suppression of the holon occupation (red curve) and an enhancement of the spinon occupation (blue curve). That is the Kondo limit. In this limit, the Kondo temperature is very small because of the small holon occupation. This can be understood through as a Gutzwiller renormalization of the hybridization $V$ due to finite holon occupancy $\tilde{V} \sim V \sqrt{\avg{n_{\chi}}}$ \cite{coleman2015introduction}. Therefore, as $n_{\chi} \rightarrow 0$, the exponential in Eq.~\eqref{eq-tkAnderson} tends to 0 exponentially. 

On the other side, as $\epsilon_f < 0$, the holons dominate over the spinons and the impurity is in the strongly mixed valence regime. There, the Fermi liquid ground state is still reached, but it is one filled with Kondo singlets, which are easy to form with the large presence of holons, hence the large Kondo temperature. Finally, on Fig.~\eqref{fig:supp1mv} (b), we see the reproduced trend of Eq.~\eqref{eq-tkAnderson}, and the exponentially smaller Kondo temperature in the bare Kondo limit.

\begin{figure}[t]
    \includegraphics[width=\linewidth]{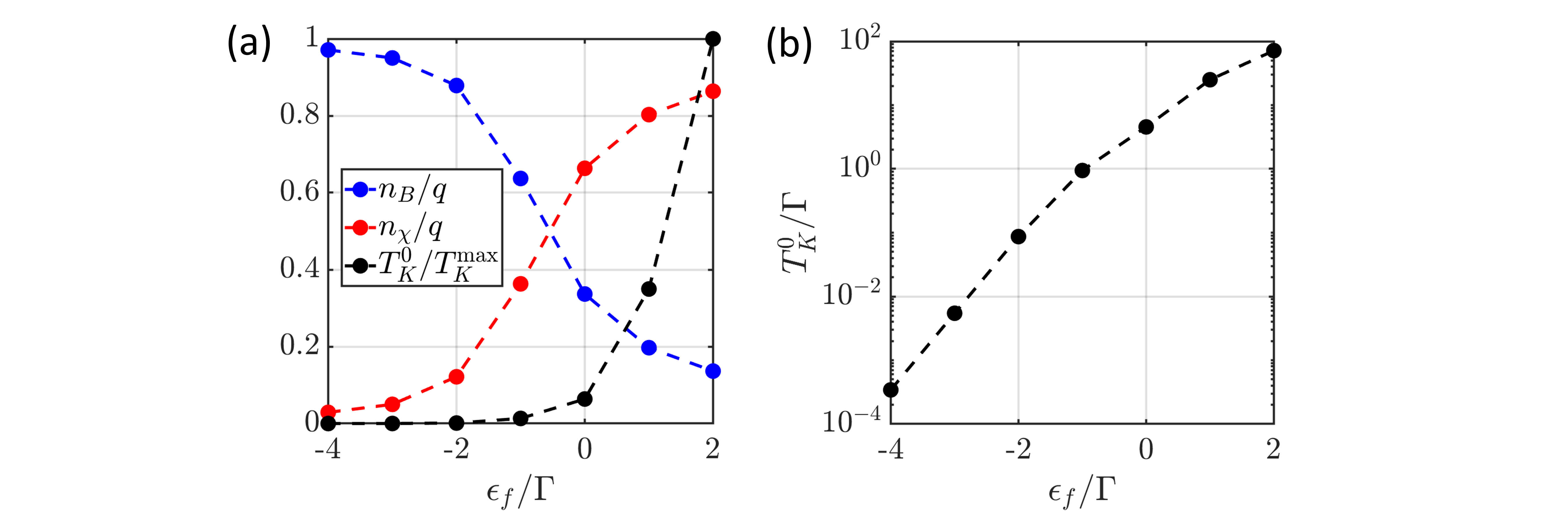}
    \caption{(a) Display of the variation in spinon and holon proportional occupation versus the holon energy level ($n_B/q$ and $n_{\chi}/q$ respectively), with $q = 2S/N$ and $n_{\xi} = - (1/\pi) \int d\omega n_{\xi}(\omega) \Imp G_{\xi} (\omega + i\delta)$. We also plot the effective Kondo temperature $T_K^0$, taken as the temperature at which the holon phase shift reaches $\delta_{\chi} > 0.98\pi$. (b) The same data for the Kondo temperature $T_K^0$, but in a logarithmic vertical axis to show the exponential spread of $T_K^0$ (following Eq.~\eqref{eq-tkAnderson}). The superscript $0$ on $T_K^0$ is to differentiate it with the effective Kondo temperature in the presence of Hund's coupling.}
    \label{fig:supp1mv}
\end{figure}

\section{II. Susceptibilities and Hubbard Operators}

Well defined susceptibilities need to be expressed in terms of the Hubbard operators. These operators, which we use to capture the contraint of no double occupancy, are defined in terms of the Schwinger bosons $b$ and holons $\chi$:

\begin{equation}
\begin{split}
X_{\alpha, a}^{(m)} = b_{m\alpha}^{\dagger} \chi_{ma} \,,\qquad & \qquad X_{a,\alpha}^{(m)} = \chi_{ma}^{\dagger} b_{m\alpha} \,,\\
X_{a,b}^{(m)} = \chi^{\dagger}_{ma} \chi_{mb} \,, \qquad & \qquad X_{\alpha, \beta}^{(m)} = b_{m\alpha}^{\dagger} b_{m\beta} \,.
\end{split}
\end{equation}

In \cite{stadler2015dynamical,stadler2019hundness, stadler2019model, walter2020uncovering, wang2020global}, the spin and orbital susceptibilities for the impurity are studied, and their behavior brings clarity to the concept of spin-orbital separation. We therefore wish to connect these observables with the large-N Schwinger boson method presented in this paper. These papers define the spin and orbital susceptibility as

\begin{equation}
\begin{split}
\chi_{\rm sp} &= \frac13 \sum_{\alpha} \bra{\widehat{S}^{\alpha}} \ket{\widehat{S}^{\alpha}}_{\omega} \,,\\
\chi_{\rm orb} &= \frac18 \sum_{a} \bra{\widehat{T}^{a}} \ket{\widehat{T}^{a}}_{\omega} \,,
\end{split}
\end{equation}

with $\widehat{T}^{a} = \frac12 \sum_{nn'\sigma} d^{\dagger}_{n \sigma} \tau^a_{nn'} d_{n'\sigma}$, $\tau^{a}$ the 8 Gell-Mann matrices such that $\text{Tr} [\tau^{a} \tau^{b}] = 2\delta_{ab}$. Similarly, the spin operators are taken as $\widehat{S}^{\alpha} = \frac12 \sum_{n\sigma \sigma'} d^{\dagger}_{n\sigma} \sigma^{\alpha}_{\sigma \sigma'} d_{n \sigma'}$, with $\text{Tr} [\sigma^{\alpha} \sigma^{\beta}] = 2\delta_{\alpha\beta}$. Their system has SU(2) spin symmetry and SU(3) orbital symmetry, hence the definition for the $\widehat{S}^{\alpha}$ and $\widehat{T}^{a}$ operators.

The expression $\bra{X} \ket{X}_{\omega}$ refers to the Fourier-transformed retarded correlation functions $-i \Theta(t) \langle [X(t), X(0)]\rangle$, with real frequency $\omega$. In order to connect with these definitions, we need to define spin and orbital operators, and then obtain the susceptibility bubbles in the large-N limit.

\subsection{Spin Susceptibility}

Consider the total spin operator on the impurity to be 

\begin{align}
    S_{\alpha \beta} (\tau) = \sum_m X_{\alpha \beta}^{(m)} (\tau) = \sum_m b^{\dagger}_{m, \alpha} (\tau) b_{m,\beta} (\tau) \,,
\end{align}

\noindent with $m$ the orbital index, and $\alpha, \beta \in [1, ..., N]$ are indices of SU(N). The total spin-spin susceptibility in imaginary time $\tau$ is then defined as 

\begin{equation}
\begin{split}
    \chi_{\rm sp} (\tau) &= \avg{\vec{S} (\tau) \cdot \vec{S} (0) } = \frac{1}{N^2} \frac13 \sum_{\alpha \beta} \avg{S_{\alpha \beta} (\tau) S_{\beta \alpha} (0) } \\
    &= \frac{1}{N^2} \frac13 \sum_{\alpha \beta} \avg{\Big[\sum_m X_{\alpha \beta}^{(m)} (\tau)\Big]\Big[\sum_{m'} X_{\beta \alpha}^{(m')} (0) \Big]} = \frac{1}{N^2} \sum_m  \sum_{\alpha \beta} \avg{X_{\alpha \beta}^{(m)} (\tau) X_{\beta \alpha}^{(0)} (0) } \;, 
\end{split}
\end{equation}

\noindent where we used orbital invariance on the $m$ index. We obtain

\begin{align}
    \chi_{\rm sp} (\tau) &= \frac{1}{N^2} \sum_m \sum_{\alpha \beta} \langle b^{\dagger}_{m,\alpha} (\tau) b_{m \beta} (\tau) b^{\dagger}_{0 \beta}(0) b_{0,\alpha}(0) \rangle \\
    &= \frac{1}{N^2} \sum_m \sum_{\alpha \beta} \Big[\cancel{\langle \wick{\c1{b}^{\dagger}_{m,\alpha} (\tau) \c1{b}_{m \beta} (\tau) \c2{b}^{\dagger}_{0 \beta}(0) \c2{b}_{0,\alpha}(0) }\rangle} + \langle \wick{\c1{b}^{\dagger}_{m,\alpha} (\tau) \c2{b}_{m \beta} (\tau) \c2{b}^{\dagger}_{0 \beta}(0) \c1{b}_{0,\alpha}(0) }\rangle \Big]\\
    &= \frac{1}{N^2} \sum_m \sum_{\alpha \beta}  \avg{ b^{\dagger}_{m,\alpha} (\tau) b_{0,\alpha}(0)} \avg{  b_{m \beta} (\tau) b^{\dagger}_{0 \beta}(0)} \\
    &= \sum_m G_B (m, \tau) G_B (- m, - \tau) \;,
\end{align}

\noindent with the definition that $G_{B} (m, \tau) = - \sum_{\alpha \beta} \langle T  b_{m \alpha} (\tau) b^{\dagger}_{0 \beta} (0) \rangle$.  The Linked-Cluster theorem allows us to reject disconnected parts (first term that is crossed), which anyway lead to a O($1/N$) contribution to the susceptibility. 

After converting to Matsubara frequencies and doing the analytic continuation $i\nu_p \rightarrow \omega + i \eta$, we obtain the finite frequency result

\begin{align}
    \chi_{\rm sp} (\omega) =  \sum_k  \int \frac{d\omega'}{2\pi}n_B (\omega') G''_B (k,\omega' +i \eta) \left[ G_B (k,\omega' - \omega - i\eta)  + G_B(k,\omega' + \omega + i\eta) \right] \;, \label{eq-spinsusc}
\end{align}

\noindent with the bosonic Green's functions $G_B (k, z) = [z - \lambda + 2\Delta \cos(k) - \Sigma_B (z)]^{-1}$ as defined in the main text. The dynamical spin susceptibility can therefore be computed as $\chi_{\rm sp}''(\omega) = -\frac{1}{\pi} \Imp{\chi_{\rm sp} (\omega)}$.

\subsection{Orbital Susceptibility}

The model we consider has $SO(3)$ symmetry, representing the angular momentum subspace of the t${}_{2g}$ ($L=1$) orbitals. The $SO(3)$ group is represented using the ${L}_{\gamma}$ operators (which are themselves linear combinations of the $(\lambda_2, \lambda_5, \lambda_7)$ Gell-Mann matrices, as SO(3) is a subgroup of SU(3)), such that $(L_{\gamma})_{mm'} = i \epsilon_{\gamma m m'}$ with $\epsilon_{\gamma m m'}$ being the usual totally antisymmetric Levi-Civita symbol. The matrices acting in the $3\times3$ orbital site space representing the angular momenta are then

\begin{align}
{L}_x =  \begin{pmatrix}
 0 &0 &0 \\
 0 &0 &-i \\
 0 &i &0  \end{pmatrix} \; , \; {L}_y =
 \begin{pmatrix}
 0 & 0 &i \\
 0 &0 & 0 \\
 -i &0 &0 
 \end{pmatrix} \; , \; {L}_z &=
 \begin{pmatrix}
 0 &-i &0 \\
 i &0 &0 \\
 0 &0 &0
 \end{pmatrix},
\end{align}

\noindent such that $[L_m, L_n] = i \epsilon_{mnk} L_k$. The angular momentum operator in the $\gamma= x, y, z$ direction, defined through the Hubbard operators, is

\begin{equation}
    \hat{L}_{\gamma} = \frac{1}{NK}\sum_{mm' \alpha a} X^{(m)}_{\alpha a} (L_{\gamma})_{ m m'} X^{(m')}_{a \alpha} = \frac{i}{NK} \sum_{mm' \alpha a} b^{\dagger}_{m,\alpha} \chi_{m,a} \epsilon_{\gamma m m'} \chi^{\dagger}_{m', a} b_{m', \alpha}  \;,
\end{equation}

\noindent with $a$ the channel index $a \in [1, \cdots, K]$, and $\alpha \in [1, \cdots, K]$ the SU(N) spin index, and $X^{(m)}_{\alpha a}$ ($X^{(m)}_{a \alpha}$) being the Hubbard operator associated with creation (annihilation) of an impurity electron at site $m$, respectively. It is clear that acting $\hat{L}_z$ on a filled state such as $\prod_{\alpha a} b^{\dagger}_{1,\alpha} \chi_{1,a} |0\rangle$ leads to its $m_z$, in this case $m_z = 1$ (one impurity electronic state is filled at $m=1$). We then define the orbital susceptibility as 

\begin{align}
    &\chi_{\rm orb} (\tau) = \frac13  \sum_{\gamma} \big[\avg{\hat{L}_{\gamma} (\tau) \hat{L}_{\gamma} (0)} - \avg{\hat{L}_{\gamma} (\tau)}\avg{\hat{L}_{\gamma} (0)} \big] \\
    &= \frac13 \frac{1}{N^2 K^2} \sum_{m m' n n '} \sum_{\alpha \beta a b} \langle X^{(m)}_{\alpha a} (\tau) X^{(m')}_{a \alpha} (\tau) X^{(n)}_{\beta b} (0) X^{(n')}_{b \beta} (0) \rangle_c \Big[ \sum_{\gamma} (L_{\gamma})_{ m m'}(L_{\gamma})_{ n n'} \Big] \\
    &= \frac13 \sum_{m m' n n '}  \Big[ \sum_{\gamma} (L_{\gamma})_{ m m'}(L_{\gamma})_{ n n'} \Big] \mathcal{A}_{m m'}^{n n'} (\tau) \;,
\end{align}

\noindent where the notation $\avg{\cdots}_c$ denotes only connected diagrams. We have the following definition for the tensor $\mathcal{A}_{m m'}^{n n'}$:

\begin{align}
    \mathcal{A}_{m m'}^{n n'} (\tau) = \frac{1}{N^2 K^2}\sum_{\alpha \beta a b} \langle \mathcal{T} X^{(m)}_{\alpha a} (\tau) X^{(m')}_{a \alpha} (\tau) X^{(n)}_{\beta b} (0) X^{(n')}_{b \beta} (0) \rangle_c \;. \label{eq:mathcalA1}
\end{align}

Vertices at time $\tau$ have one Hubbard operator entering and one leaving; this creates a complicated vertex for bosons and holons (see Fig.~\ref{fig:4opt} (a)). There are three types of connected diagrams one can create after Wick contracting the boson and holon lines (see Fig.~\ref{fig:4opt} (b-d)).

\begin{figure}
    \centering
    \includegraphics[width=0.8\linewidth]{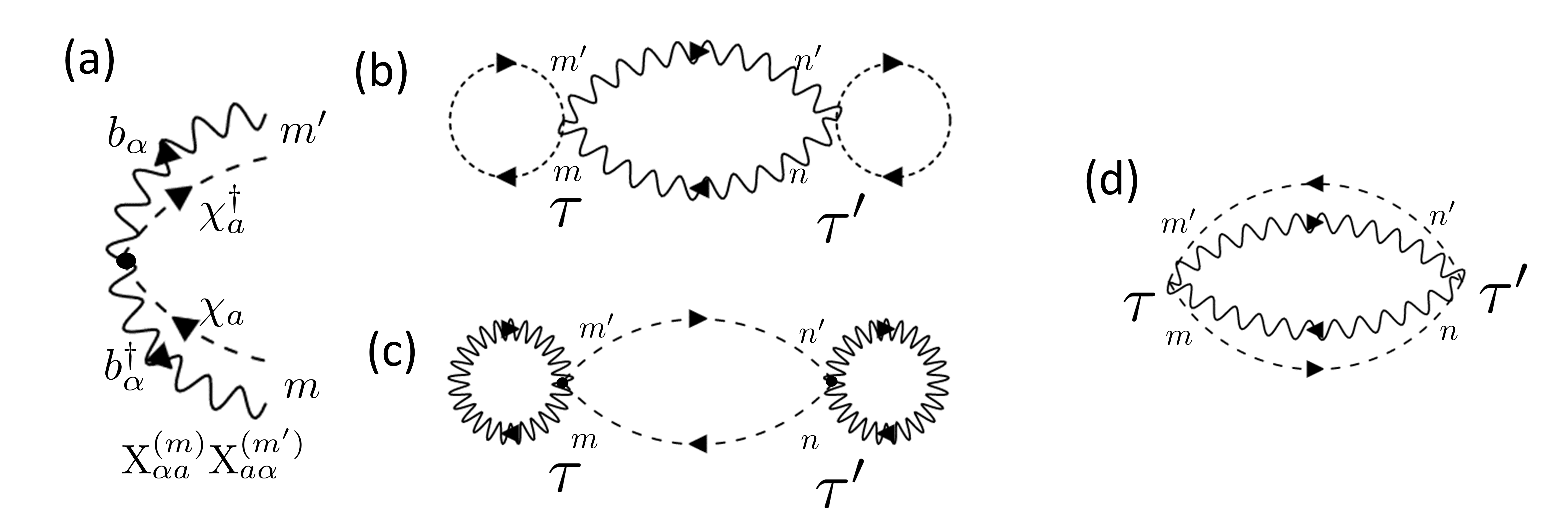}
    \caption{(a) The vertex for the Hubbard operators $\mathrm{X}_{\alpha a}^{(m)}$ for sites $m$ and $m'$. Dashed lines are holon propagators $G_{\chi}$ and wavy lines are spinon propagators $G_b$. (b-d) The three connected Wick contractions that can be created from the vertex presented in (a). Each closed dashed loop carries a summation over $a$ indices ($O(K)$ sum) while each closed wavy loop carries a summation over $\alpha$ indices ($O(N)$ sum). Whereas (b) and (c) scale like $O(N^2 K)$ and $O(K^2 N)$ respectively, (d) scales like $O(NK)$, and we discard it in the large-N limit.}
    \label{fig:4opt}
\end{figure}

Thus, the calculation of the susceptibilities includes objects (b-d) from figure~\ref{fig:4opt}. (b-c) respectively lead to $O(1/K)$ and $O(1/N)$ contributions to the susceptibility, while (d) is $O(1/NK)$, as seen by summing repeated indices for each bosonic and fermionic loop. The expression for the sum of these three connected graphs is therefore, in the most general case,

\begin{align}
    \mathcal{A}_{m m'}^{n n'} (\tau) &= \frac{1}{N^2 K^2}\sum_{\alpha \beta a b} \langle \mathcal{T} X^{(m)}_{\alpha a} (\tau) X^{(m')}_{a \alpha} (\tau) X^{(n)}_{\beta b} (0) X^{(n')}_{b \beta} (0) \rangle_c \\
    & = \frac{1}{N^2 K^2} \sum_{\alpha \beta a b} \langle \mathcal{T} b^{\dagger}_{m,\alpha} (\tau) \chi_{m,a} (\tau) \chi^{\dagger}_{m' a} (\tau) b_{m' \alpha} (\tau) b^{\dagger}_{n \beta}(0) \chi_{n b}(0) \chi^{\dagger}_{n',b}(0) b_{n',\beta}(0)  \rangle_c \\
    \begin{split}
    &=  \frac{1}{N^2 K^2}\sum_{\alpha \beta a b} \langle \mathcal{T} \wick{ \c1{b}^{\dagger}_{m,\alpha} (\tau) \c1{b}_{m' \alpha} (\tau) \c2{b}^{\dagger}_{n \beta}(0) \c2{b}_{n',\beta}(0)  \c3{\chi}_{m,a} (\tau) \c4{\chi}^{\dagger}_{m' a} (\tau) \c4{\chi}_{n b}(0) \c3{\chi}^{\dagger}_{n',b}(0) }\rangle  \\
    &+  \frac{1}{N^2 K^2}\sum_{\alpha \beta a b} \langle \mathcal{T} \wick{ \c1{b}^{\dagger}_{m,\alpha} (\tau) \c2{b}_{m' \alpha} (\tau) \c2{b}^{\dagger}_{n \beta}(0) \c1{b}_{n',\beta}(0)  \c3{\chi}_{m,a} (\tau) \c3{\chi}^{\dagger}_{m' a} (\tau) \c4{\chi}_{n b}(0) \c4{\chi}^{\dagger}_{n',b}(0) }\rangle  \\
    &+\frac{1}{N^2 K^2}\sum_{\alpha \beta a b} \langle \mathcal{T} \wick{ \c1{b}^{\dagger}_{m,\alpha} (\tau) \c2{b}_{m' \alpha} (\tau) \c2{b}^{\dagger}_{n \beta}(0) \c1{b}_{n',\beta}(0)  \c3{\chi}_{m,a} (\tau) \c4{\chi}^{\dagger}_{m' a} (\tau) \c4{\chi}_{n b}(0) \c3{\chi}^{\dagger}_{n',b}(0) }\rangle 
    \end{split} \label{eqS20} \\
    \begin{split} 
    &=  -\frac{1}{K}G_B (m', m, 0^-)  G_B (n', n, 0^-)   G_{\chi} (m, n', \tau) G_{\chi} (n, m', - \tau)  \\
    &+ \frac{1}{N} G_B (n', m, -\tau) G_B (m', n, \tau)  G_{\chi} (m, m', 0^+)  G_{\chi} (n, n', 0^+)   \\
    & -\frac{1}{NK} G_B (n', m, -\tau) G_B (m', n, \tau) G_{\chi} (m, n', \tau) G_{\chi} (n, m', - \tau) \;,
    \end{split} \label{eqS21}
\end{align}

\noindent where the $1/N$ and $1/K$ factors come from $G_B = \frac{1}{N} \sum_{\alpha} G_{B, \alpha} (\tau) =  - \frac{1}{N}\sum_{\alpha} \langle \mathcal{T} b_{\alpha} (\tau) b^{\dagger}_{\alpha} (0) \rangle$ and $G_{\alpha \beta} = \delta_{\alpha \beta} G_{\alpha}$. For Eqs.~\eqref{eqS20} and \eqref{eqS21}, the three expressions correspond respectively to the diagrams (b) (c) and (d) from figure~\ref{fig:4opt}. Furthermore, we are using the notation that

\begin{equation}
    G (m, n, \tau) = - \langle T \mathcal{O}_{m} (\tau)  \mathcal{O}_{n}^{\dagger} (0) \rangle  \;,
\end{equation}

\noindent with $\mathcal{O}$ a bosonic or fermionic operator. Returning to the expression for the orbital susceptibility, we have that

\begin{equation}
\chi_{\rm orb}^{\gamma} (\tau)  = \frac13 \sum_{m m' n n '} (L_{\gamma})_{ m m'}(L_{\gamma})_{ n n'} \mathcal{A}_{m m'}^{n n'} (\tau) \;,
\end{equation}

\noindent such that we obtain the following expressions in the three orbitals $\gamma = x,y,z$:

\begin{align}
    \chi_{\rm orb}^x (\tau) &= \frac13  \Big[ -\mathcal{A}_{3,2}^{3,2} (\tau)  -\mathcal{A}_{2,3}^{2,3} (\tau) + \mathcal{A}_{3,2}^{2,3} (\tau) + \mathcal{A}_{2,3}^{3,2} (\tau)\Big] \;,\\
    \chi_{\rm orb}^y (\tau) &= \frac13  \Big[ -\mathcal{A}_{3,1}^{3,1} (\tau)  -\mathcal{A}_{1,3}^{1,3} (\tau) + \mathcal{A}_{3,1}^{1,3} (\tau) + \mathcal{A}_{1,3}^{3,1} (\tau)\Big] \;,\\
    \chi_{\rm orb}^z (\tau) &= \frac13  \Big[ -\mathcal{A}_{1,2}^{1,2} (\tau)  -\mathcal{A}_{2,1}^{2,1} (\tau) + \mathcal{A}_{1,2}^{2,1} (\tau) + \mathcal{A}_{2,1}^{1,2} (\tau)\Big] \;.
\end{align}

In the case of $\chi_{\rm orb}^x (\tau)$, we therefore need to evaluate $\mathcal{A}_{3,2}^{3,2} (\tau)$ and $\mathcal{A}_{3,2}^{2,3} (\tau)$. To do so, we first remark that since the term in the action for the holons is diagonal in orbital space ($S_{\chi} \propto \epsilon_f \sum_m \chi^{\dagger}_{ma} \chi_{ma}$), the holon Green's function is diagonal in orbital space. Hence,

\begin{equation}
    G_{\chi} (x, y, \tau) = - \frac{1}{K} \sum_a \langle \mathcal{T} \chi_{x, a} (\tau) \chi^{\dagger}_{y, a} (0) \rangle  = \delta_{xy} G_{\chi}(\tau).
\end{equation}

Once we feed those results into the expression for $\mathcal{A}_{m m'}^{n n'} (\tau)$, we obtain

\begin{equation}
    \begin{split}
    \mathcal{A}_{3,2}^{3,2} (\tau) &=  -\frac{1}{K} G_B (2,3, 0^-)  G_B (2,3, 0^-)   \cancel{G_{\chi} (3, 2, \tau) G_{\chi} (2, 3, - \tau)}  \\
    &+ \frac{1}{N} G_B (2, 3, -\tau) G_B (2, 3, \tau)  \cancel{G_{\chi} (3 - 2, 0^+)  G_{\chi} (3, 2, 0^+)}  \\
    & -\frac{1}{NK} G_B (2, 3, -\tau) G_B (2, 3, \tau) \cancel{G_{\chi} (3, 2, \tau) G_{\chi} (3, 2, - \tau)} \\
    &= 0 =  \mathcal{A}_{2,3}^{2,3} (\tau) \;,
\end{split}
\end{equation}

\noindent where $G_{\chi}(3,2,\tau) = 0$ because there is no orbital mixing of the holons, while

\begin{equation}
    \begin{split}
    \mathcal{A}_{3, 2}^{2,3} (\tau) &=  -\frac{1}{K}G_B (2,3, 0^-)  G_B (3, 2, 0^-)   G_{\chi} (3, 3, \tau) G_{\chi} (2, 2, - \tau)  \\
    &+ \frac{1}{N} G_B (3, 3, -\tau) G_B (2, 2, \tau)  \cancel{G_{\chi} (3, 2, 0^+)  G_{\chi} (2, 3, 0^+)}  \\
    & -\frac{1}{NK} G_B (3, 3, -\tau) G_B (2, 2, \tau) G_{\chi} (3, 3, \tau) G_{\chi} (2, 2, - \tau) \\
    &= -\frac{1}{K} G_{\chi} (\tau) G_{\chi} (- \tau) \Big[ G_B (2,3, 0^-)  G_B (3, 2, 0^-) + \frac{1}{N} G_B (3, 3, -\tau) G_B (2, 2, \tau) \Big] \;.
\end{split}
\end{equation}

In the limit of large-N, only the first contribution will be relevant. The other contributions from the $y$ and $z$ sectors will lead to essentially the same results, with a change of index. Therefore, we identify 

\begin{equation}
    \chi_{\rm orb} (\tau) = \sum_{\gamma} \chi_{\rm orb}^{\gamma} =  \frac{1}{K} \Big( \frac{\Delta}{J_H} \Big)^2 G_{\chi} (\tau) G_{\chi} (- \tau) \;.
\end{equation}

The constant factor of $\Big( \frac{\Delta}{J_H} \Big)^2$ in this expression can be obtained after Fourier transforming the bosonic Green's function:

\begin{equation}
    \frac13 \sum_{m} G_B (m,m+1, 0^-)G_B (m+1, m, 0^-) = [G_B (k=0, 0^-) - G_B (k=\pm 2\pi/3, 0^-)]^2 \;,
\end{equation}

\noindent where $G_B (k, z) = [z - \lambda - 2\Delta \cos(k) - \Sigma_B (z)]^{-1}$. We have that 

\begin{equation}
    G_B (k=0, 0^-) - G_B (k=\pm 2\pi/3, 0^-) = \sum_k \int_{-\infty}^{\infty}  \frac{d\omega}{\pi} n_B (\omega) \text{Im} [ \cos(k) G_B (k, \omega + i \eta)] = \frac{\Delta}{J_H}\;.
\end{equation}

This is the same definition as the constraint on the Schwinger bosons, which leads to the $\Delta/J_H$ factor, and hence completes the expression. 

After converting to Matsubara frequencies and doing the analytic continuation $i\omega_p \rightarrow \omega + i \eta$, we obtain the finite frequency result

\begin{align}
    \chi_{\rm orb} (\omega) = \frac{1}{K} \Big( \frac{\Delta}{J_H} \Big)^2  \int \frac{d\omega'}{2\pi}n_F (\omega') G''_{\chi} (\omega' +i \eta) \left[ G_{\chi} (\omega' - \omega - i\eta)  + G_{\chi}(\omega' + \omega + i\eta) \right]\;,
\end{align}

\noindent with the holonic Green's functions $G_{\chi} (z) = [z - \lambda - \epsilon_f - \Sigma_{\chi} (z)]^{-1}$ as defined in the main text. The dynamical orbital susceptibility can therefore be computed as $\chi_{\rm orb}''(\omega) = -\frac{1}{\pi} \Imp{\chi_{\rm orb} (\omega)}$. This concludes the derivation of the expressions presented in Eq.~(7) and Fig.~3.

\section{III. Details on the Single Iteration Approach}

In this section, we analytically evaluate a single self-energy correction to the holon's propagator, therefore obtaining the logarithmically dependent effective Kondo coupling, as mentioned in Eq.~\eqref{eq-jkeff0} in the main text. We consider the single loop correction to the holon propagator, 

\begin{align}
     G_{\chi, 0}^{-1} (z) &= z - \epsilon_f - \lambda \; , \\
    G_{\chi, 1}^{-1} (z) &= G_{\chi, 0}^{-1} (z) - V^2 \Sigma_{\chi,0} (z)\;. 
\end{align}

\noindent In a previous approach \cite{drouin2021emergent}, the evaluation of the first order self-energy at $z = 0 + i\delta$ led to an analytical expression for the effective Kondo temperature in presence of Hund's coupling. The argument there was that the holons have a pole that crosses from $\omega >0$ to $\omega <0$ at $T = T_K^{\rm eff}$. Here, we expand this work to include the frequency dependence of the self-energy, based on other works on underscreened Kondo models \cite{coleman2003singular, coleman2005quantum}. 

A single iteration is conducted by estimating $\Sigma_{\chi,0}$ in a single loop, i.e. through the bare propagators for the spinons and conduction electrons. In imaginary time, this is written as

\begin{equation}
    \Sigma_{\chi,0} (\tau) = g_{c,0} (-\tau) G_{B,0}(\tau) \;,
\end{equation}

\noindent where we use the following bare propagators in the presence of Hund's coupling.

\begin{align}
    g_{c,0}(z) = \int d\epsilon \frac{\rho(\epsilon)}{z-\epsilon} \qquad ,\qquad G_{B,0}(z) = \frac{1}{3} \Big[ (z - \lambda')^{-1} + 2(z - \lambda' - 3\Delta)^{-1} \Big]\;.
\end{align}

Passing from imaginary time to Matsubara frequencies leads to 

\begin{equation}
    \Sigma_{\chi,0} (i \omega_m) = \frac{1}{3} \int d\epsilon \rho(\epsilon) \sum_n \frac{1}{i\omega_n - \epsilon} \Big[ \frac{1}{i\omega_n + i\omega_m - \lambda'}+  \frac{2}{i\omega_n + i\omega_m - \lambda' - 3\Delta} \Big]\;,
\end{equation}

\noindent which, after doing the Matsubara sum over fermionic frequencies, leads to

\begin{equation}
    \Sigma_{\chi,0} (i \omega_m) = \frac{1}{3} \int d\epsilon \rho(\epsilon) \Big[\frac{n_F (\epsilon_k) + n_B(\lambda')}{\epsilon - (\lambda' - i \omega_n)} + 2 \frac{n_F (\epsilon_k) + n_B(\lambda' + 3\Delta)}{\epsilon - (\lambda'+3\Delta - i \omega_n)} \Big]\;.
\end{equation}

We can then analytically continue this expression $i\omega_n \rightarrow \omega + i\delta$. Finally, progress can be made to evaluate analytically this integral. Using $\rho(\epsilon) = \rho \Phi(\epsilon)$ with $\rho = 1/2D$, such that $\Phi(\epsilon) D^2/(\omega^2 + D^2)$ for the conduction electrons' density of states ($\Phi$ acts as a soft cutoff at $\omega = D$, the bandwidth) simplifies the problem. Integrals of this sort are regularly found in Kondo problems \cite{coleman2015introduction}, and can be evaluated through the use of the digamma function $\Psi(z) = d\ln \Gamma(z) /dz$. The general result we use is that 

\begin{align}
    \int d \omega' \Phi(\omega') \frac{ n_F (\omega')}{\omega' - A}
    &=  \left[ \Psi \left(\frac{1}{2} - \frac{A}{2\pi i T}\right) - \ln \left( \frac{D}{2\pi T}\right) \right] - \frac{\pi }{2} \frac{D}{iD - A}\;.
\end{align}

Therefore, the full result is obtained:

\begin{equation}
\begin{split}
    \Sigma_{\chi,0} (\omega) &= -\frac{\rho}{3} \Big(1 + \frac{2}{1 + \frac{\Delta^2}{D^2}} \Big)\ln \Big( \frac{D}{2\pi T} \Big) + \frac{\rho}{3} \Big[ \Psi \Big(\frac{1}{2} + \frac{\lambda' - \omega}{2\pi i T}\Big) + \frac{2}{1 + \frac{\Delta^2}{D^2}} \Psi \Big(\frac{1}{2} + \frac{\lambda' + 3 \Delta - \omega}{2\pi i T}\Big) \Big] \\
    &+ \frac{\pi \rho}{6} [\coth (-\beta \lambda'/2)  + 2\coth (-\beta (\lambda' + 3\Delta)/2)  ] \, ,
\end{split}
\end{equation}

\noindent and the single iteration result for the holon's Green's function is then 

\begin{equation}
\begin{split}
    G_{\chi, 1}^{-1} (\omega) &= \omega - \epsilon_f - \lambda' + 2 \Delta \\
    &+\frac{\rho V^2}{3} \Big(1 + \frac{2}{1 + \frac{\Delta^2}{D^2}} \Big)\ln \Big( \frac{D}{2\pi T} \Big) - \frac{\rho V^2}{3} \Big[ \Psi \Big(\frac{1}{2} + \frac{\lambda' - \omega}{2\pi i T}\Big) + \frac{2}{1 + \frac{\Delta^2}{D^2}} \Psi \Big(\frac{1}{2} + \frac{\lambda' + 3 \Delta - \omega}{2\pi i T}\Big) \Big] \\
    & - \frac{\pi \rho V^2}{6} [\coth (-\beta \lambda'/2)  + 2\coth (-\beta (\lambda' + 3\Delta)/2)  ] \,, \label{singleithG}
\end{split}
\end{equation}

\noindent where $\lambda' = \lambda + 2\Delta$ is the chemical potential for the Schwinger bosons in the aligned state ($k=0$). The unaligned state is gapped. The value of $\lambda'$ is obtained by satisfying the constraint concerning the number of bosons. Quite generally, we will have that $n_B(\lambda) = q/\alpha$ with $\alpha > 1$ where $q = 2S/N$. $\alpha = 1$ corresponds to no Hund's coupling and no holons ($n_{\chi} = 0$). The presence of holons reduces the $T=0$ spinon occupation and increases $\alpha$. Therefore, as $T \rightarrow 0$, $\lambda' \rightarrow T \ln(1 + \alpha/q)$. 

Before exploring the explicit frequency dependence of the holon's Green's function, it is very useful to extract the universal temperature scale $T_K^{\rm eff}$. The Kondo temperature, signaling the advent of the Nozières Fermi liquid at $T<T_K^{\rm eff}$, is intimately related to a shift of the pole $G_{\chi}$. As it was used before \cite{drouin2021emergent}, a useful tool to estimate $T_K^{\rm eff}$ is that 

\begin{equation}
    \Rep G_{\chi}^{-1}(z) \vert_{z=0+i\delta} = 0\;, \qquad \text{at} \qquad T=T_K^{\rm eff} \;.\label{tkeffcrit}
\end{equation}

In extracting the effective Kondo temperature, we are mainly concerned with two limits. Firstly, in the absence of Hund's coupling, we can set $\Delta = 0$. This is a conventional Anderson impurity problem, and an effective Kondo coupling $J_K \sim \frac{V^2}{\epsilon_f + \lambda}$ can be obtained by a Schrieffer Wolff transformation. The same $J_K$ scale emerges out of this method. In this case, the holon's Green's function is simply

\begin{align}
    G_{\chi, 1}^{-1} (\omega) &= \omega - \epsilon_f - \lambda'+\rho V^2 \ln \Big( \frac{D}{2\pi T} \Big) - \rho V^2 \Psi \Big(\frac{1}{2} + \frac{\lambda' - \omega}{2\pi i T}\Big)  - \frac{\pi \rho V^2}{2} \coth (-\beta \lambda'/2) \,.
\end{align}

Therefore, the Kondo temperature criteria of Eq.~\eqref{tkeffcrit} leads to the following equation

\begin{align}
    \frac{\epsilon_f + \lambda'}{\rho V^2} &= \ln \Big( \frac{D}{2\pi T_K^{\rm eff}} \Big) - \Rep \Psi \Big(\frac{1}{2} + \frac{\lambda'}{2\pi i T_K^{\rm eff}}\Big)  - \frac{\pi}{2} \coth (-\beta \lambda'/2) \,. \label{tkeffnoHund}
\end{align}

Quite generally, we will have that $n_B(\lambda) = q/\alpha$ with $\alpha > 1$ where $q = 2S/N$. $\alpha = 1$ corresponds to no Hund's coupling and no holons ($n_{\chi} = 0$). The presence of holons reduces the $T=0$ spinon occupation and increases $\alpha$. Therefore, we have $\lambda \rightarrow T \ln(1 + \alpha/q)$. This simplifies the digamma term as $\ln[\ln(1 + \alpha/q)/2\pi]$. Furthermore, we approximate the digamma function such that $\Rep \Psi(z) \simeq \ln(|z|)$ and simplifying Eq.~\eqref{tkeffnoHund}, we get

\begin{align}
\frac{T_K^{\rm eff}}{D} &= \mathcal{F}(q) \exp{\Big[- \frac{\epsilon_f + \lambda}{\rho V^2}\Big]} \;, \label{eq-tkeffsuppSW}\\
\mathcal{F}(q) &=\frac{\exp{[-\frac{\pi}{2} \coth (-\ln(1 + \alpha/q)/2)]}}{\ln(1 + \alpha/q)} \;,
\end{align}

\noindent where we recover an analogous form to the typical Kondo temperature $T_K/D \simeq \exp{[-1/\rho J_K]}$ with $J_K \sim \frac{V^2}{\epsilon_f + \lambda}$ as expected from a Schrieffer Wolff transformation. 

In the case of large Hund's coupling, the spinon gap can be taken to be larger than the electronic bandwidth ($\Delta \gg D$), and all spinons will be in the aligned ground state. Furthermore, the chemical potential adjusts itself so that $\lambda' = \lambda -2\Delta \simeq T \ln(1 + \alpha/q)$. In this case, the holon's Green's function is 

\begin{align}
    G_{\chi, 1}^{-1} (\omega) &= \omega - \epsilon_f - \lambda  +\frac{\rho V^2}{3} \ln \Big( \frac{D}{2\pi T} \Big) - \frac{\rho V^2}{3} \Psi \Big(\frac{1}{2} + \frac{\lambda' - \omega}{2\pi i T}\Big) - \frac{\pi \rho V^2}{6} [\coth (-\beta \lambda'/2)  - 2 ] \,. \label{eq-48}
\end{align}

Therefore, the Kondo temperature criteria of Eq.~\eqref{tkeffcrit} leads to 

\begin{align}
     3 \frac{\epsilon_f + \lambda}{\rho V^2} &=  \ln \Big( \frac{D}{2\pi T_{K}^{\rm eff}} \Big) - \Rep \Psi \Big(\frac{1}{2} + \frac{\ln(1 + \alpha/q)}{2\pi i}\Big) - \frac{\pi}{2} [\coth (-\ln(1 + \alpha/q)/2)  - 2 ] \,,
\end{align}

\noindent and solving for $T_K^{\rm eff}$ leads to

\begin{align}
\frac{T_K^{\rm eff}}{D} &= \mathcal{K}(q) \exp{\Big[- 3\frac{\epsilon_f + \lambda}{\rho V^2}\Big]} \;,\\
\mathcal{K}(q) &=\frac{\exp{[-\frac{\pi}{2} \coth (-\ln(1 + \alpha/q)/2) + \pi]}}{\ln(1 + \alpha/q)}\;,
\end{align}

\noindent which is equivalent, generically, to the following form

\begin{align}
\frac{T_K^{\rm eff, H} }{D} &\propto \Big( \frac{T_K^{\rm eff,0} }{D} \Big)^3 \;, \\
\frac{T_K^{\rm eff,0} }{D} &\propto  \exp{\Big[- \frac{\epsilon_f + \lambda}{\rho V^2}\Big]}\;,
\end{align}

\noindent with the proportionality factor being a function of $q$, comprising the functions $\mathcal{F}(q)$ and $\mathcal{K}(q)$. This is a simple recovery of both the effective Kondo temperature for an Anderson model \cite{coleman2015introduction}, as well as the Schrieffer effect for strong Hund's coupling \cite{schrieffer1967kondo}. Therefore, we will generically associate 

\begin{align}
    \frac{\epsilon_f + \lambda}{\rho V^2} = \frac{\mathcal{C}}{n} \ln \Big( \frac{D}{2\pi T_K^{\rm eff}}\Big) \;, \label{eqEfKondo}
\end{align}

\noindent with $n = 1$ for $J_H = 0$ (no Hund's coupling) and $n=3$ for $J_H \gg D$ (strong Hund's coupling). $\mathcal{C}$ is a $O(1)$ number depending on $\alpha/q$.

We are now fully able to study the holon Green's function's full frequency dependence of Eq.~\ref{eq-48}.  We consider frequencies $\omega > \max \{ T, T_K^{\rm eff} \}$. In that case, the digamma functions in Eq.~\eqref{eq-48} can be approximated as $\Psi(a + \omega) \simeq \ln(\omega)$ for large arguments $\omega$. Furthermore, we take advantage of the relation of Eq.~\eqref{eqEfKondo}, and are left with  

\begin{equation}
\begin{split}
    G_{\chi, 1}^{-1} (\omega) &= \omega - \rho V^2 \Big[ \frac{ \mathcal{C}}{n} \ln \Big( \frac{D}{2 T_K^{\rm eff}}\Big) -\frac{1}{n} \ln \Big( \frac{D}{2\pi T} \Big) + \frac{1}{n} \ln \Big(\frac{\omega}{2\pi T}\Big) - \mathcal{D} \Big] \,, \label{eqlogomega}
\end{split}
\end{equation}

\noindent for the general holon Green's function at any temperature. Simplifying the logarithms, one finds that the holon's Green's function for the mixed valence problem, once the self-energy has been taken into account, is 

\begin{equation}
\begin{split}
    G_{\chi, 1}^{-1} (\omega) &= \omega -  \frac{\Gamma}{\pi n} \ln \Big(\frac{\omega}{ T_{K}^{\rm eff}}\Big)  + \mathcal{B} \,. \label{eqlogomega2}
\end{split}
\end{equation}

In the cases studied in this paper, the effective Kondo temperature is very small due to both the presence of holons in the system ($T_K^{0}$ is small), and because of strong Hund's coupling ($n=3$). Note that, for $n=3$, this is only valid for $\omega < T_{\rm orb}$, as we have "locked" the spins at the different orbitals together when putting $n=3$. 

\subsection{Mapping to a Kondo problem}

Upon inspection, we can see from Eq.~\ref{eqlogomega2} that if $T_K^{\rm eff} < \omega < \min \{ \Gamma, T_{\rm orb} \}$, then the expression is dominated by the second term. This factor $\mathcal{D} = \min \{ \Gamma, T_{\rm orb} \}$ is a natural high-energy cutoff for the regime of validity of our expressions. For these small frequencies yet above $T_K^{\rm eff}$, the logarithm part due to the $\omega$-dependent Kondo coupling dominates the Green's function. 

In this regime, we find that the physical holons $\chi^{\dagger}$ can be replaced by simple Grassmann fields $\bar{\chi} = V \chi^{\dagger}$, where we have absorbed the hybridization $V$ into their definition. This is to connect to the Parcollet-Georges decoupling \cite{parcollet1997transition, parcollet1998overscreened, coleman2005sum, rech2006schwinger} of the Kondo interaction through the use of Grassmann variables, as we have used in Ref. \cite{komijani2018model, drouin2021emergent}. In the infinite-U Anderson model, we have that the holon part of the action is [see Eq.~\eqref{eq-model}]

\begin{equation}
    H_{\rm mv} = \sum_m V( c^\dagger_{0 m \alpha a} \chi^{\dagger}_{m,a} b_{m\alpha} + h.c.) + (\epsilon_f + \lambda) \chi^{\dagger}_{m,a} \chi_{m,a} \;,
\end{equation}

\noindent where the effect of Eq.~\eqref{eqlogomega2} is to renormalize the holon energy level from $\epsilon_f + \lambda$ to $\epsilon^{\ast} (\omega) =  \frac{\rho V^2}{n} \ln \Big(\frac{\omega}{ T_{K}^{\rm eff}}\Big)$. In the Kondo model \cite{drouin2021emergent}, we had

\begin{equation}
    H_{\rm kondo} = \sum_m ( c^\dagger_{0 m \alpha a} \bar{\chi}_{m,a} b_{m\alpha} + h.c.) + \frac{\bar{\chi}_{m,a} \chi_{m,a}}{J_K} \;.
\end{equation}

We can see that one can connect the two Hamiltonians with the association $\bar{\chi} = V \chi^{\dagger}$. This leads to an effective Kondo coupling

\begin{equation}
    J_K^{\rm eff} \simeq \frac{V^2}{\epsilon^{\ast} (\omega)} = \frac{n}{\rho \ln \Big(\frac{\omega}{ T_{K}^{\rm eff}}\Big)} \,. \label{eqlogomega3}
\end{equation}

The new Kondo Green's function for our holons will be noted as $\tilde{G}_{\chi}(\omega) = -\avg{T \chi(\tau) \bar{\chi}(0)}$, in terms of the newly introduced Grassmann quantities. Because of the absorption of a factor of $V$ into the definition of $\bar{\chi}$, we get  $V^2 \tilde{G}_{\chi}(\omega) = G_{\chi,1} (\omega)$. There is no dynamics (no $\partial_{\tau}$ term) in the action for these Grassmannian variables. Therefore, the $\omega$ in Eq.~\eqref{eqlogomega2} can be removed in the expression for $\tilde{G}_{\chi}$. The final Kondo Green's function is  

\begin{equation}
\begin{split}
    \tilde{G}_{\chi} (\omega) &\simeq \Big[ - \frac{1}{J_K^{\rm eff}} \Big]^{-1}\;.
\end{split}
\end{equation}

This replacement is only valid in the limit where $\omega/T_K^{\rm eff} > 1$ while $\max (T, \omega) < J_H$. In this regime, the nearly frozen charge degrees of freedom means that we can neglect the holon's dynamics and get a very simple instantaneous effective imaginary time Green's function

\begin{equation}
\begin{split}
    \tilde{G}_{\chi} (\tau) &= - J_K^{\rm eff} \delta(\tau)\;. \label{gchi-imtime}
\end{split}
\end{equation}

\noindent This concludes the derivation of Eq.~\eqref{eq-jkeff0} in the main text. This result is further used in the next section for second order contribution to the dynamical spin susceptibility. For large frequencies $\omega = D$ one recovers the bare $J_K^{\rm eff}$ value of Eq.~\eqref{eq-tkeffsuppSW}.

\section{IV. Logarithmic Correction to the Spin Susceptibility}

In this section, we consider whether the combination of the single iteration holon propagator, as derived above in Eq.~\eqref{eqlogomega3}, with the spinon susceptibility bubble, is enough to obtain the correct scaling form for the dynamical spin susceptibility, as seen if Fig.~\ref{fig:susceptibility}. Former insight on this is provided by Ref.~\cite{koller2005singular}, in which a running Kondo coupling is introduced in the context of underscreened Kondo models such that $\tilde{J} (\omega) = 1/\ln(\omega/T_0)$ with $T_0 \sim T_K$ the onset on underscreened features. Through a second order correction to the impurity spin dynamics as electrons are scattered on the impurity through the Kondo coupling $J_K$, this running coupling is then used to predict the spin susceptibility behavior of $\chi''(\omega) \sim [\omega \ln^2 (\omega/T_K)]^{-1}$. 

The question then becomes whether we obtain such a correction to $\chi_{\rm sp}$ in this Schwinger boson formalism too. The previous section showed that a single iteration of the self-energy equations leads to a running effective Kondo coupling as the holon propagator. This section focuses on the use of this propagator to calculate the second order correction to spinon dynamics as they scatter into holons and conduction electrons.

\begin{figure}
    \centering
    \includegraphics[width=0.9\linewidth]{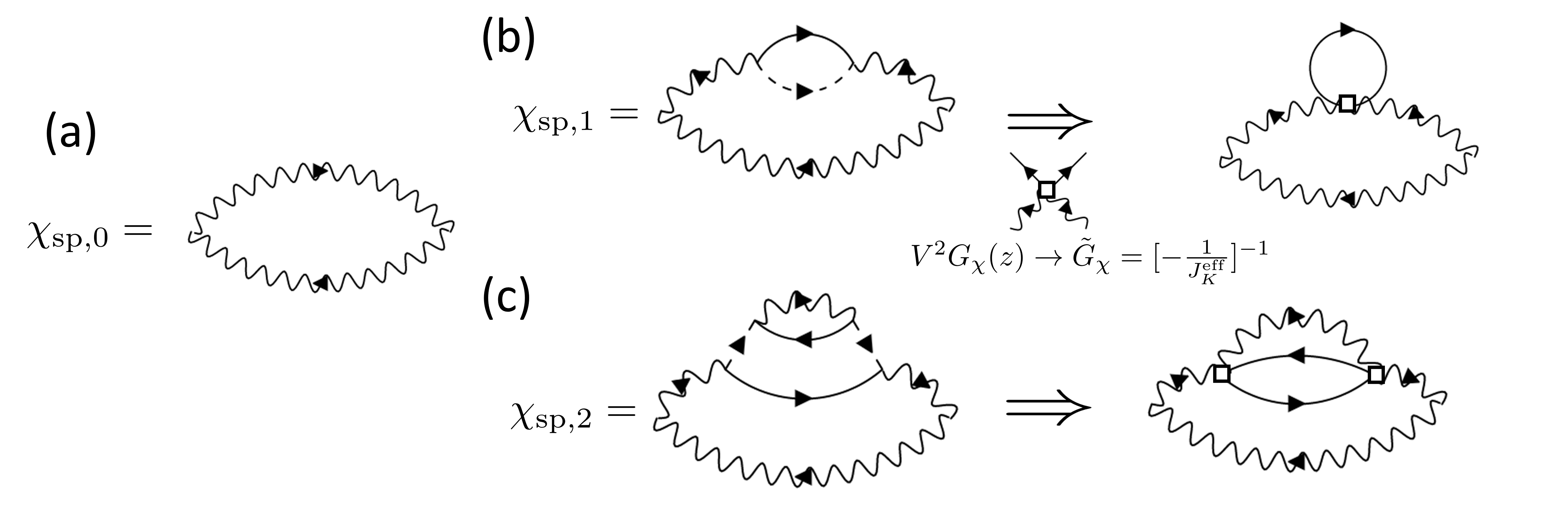}
    \caption{(a) Full spinon susceptibility with the dressed propagators.  (b) Zeroth order contribution to the spin susceptibility where only the bare spinon propagrators are taken into account. (b) First order scattering of a spinon into the conduction sea through the Kondo effect. In the limit of $T_K^{\rm eff} \ll \omega \ll T_{\rm orb}$, due to strong Hund's coupling, the holon propagator can be replaced by the Kondo-like instantaneous propagator. This diagram then has a bubble correction from the conduction electrons. (c) Second order correction to the spinon susceptibility. Taking the same limit of an instantaneous holon propagator, we recover the rightmost diagram of (c), which is calculated explicitly in Eq.~\eqref{eq-imsuscSP4} and is responsible for the logarithmic features in the dynamical spin susceptibility.}
    \label{fig:2orders}
\end{figure}

We first consider the zeroth order contribution to the spin susceptibility. We use the result of Eq.~\eqref{eq-spinsusc}, and replace $G_B (k, z) \rightarrow G_{B,0} (k,z)$. Furthermore, we can consider both cases of strong Hund's coupling and no Hund's coupling in the same breath. In the case $J_H = 0$, $G_{B,0} (k,z) = G_{B,0}(z) = [z - \lambda]^{-1}$. The chemical potential $\lambda$ is found by satisfying the contraint on boson number, i.e. we have that $n_B(\lambda) \simeq q = 2S/N$. 

In the case of finite $J_K$, the bare spinon propagators are $G_{B,0} (k=0, z) = [z - \lambda + 2\Delta]^{-1}$ and $G_{B,0} (k=\pm 2\pi/3, z) = [z - \lambda - \Delta]^{-1}$. The chemical potential adjusts itself to the gap in the spinon dispersion, and we have that $n_B(\lambda - 2\Delta) \simeq 3q$. Noting $\lambda' = \lambda - 2\Delta$, we therefore have $G_{B,0} (k=0, z) = [z - \lambda']^{-1}$ and $G_{B,0} (k=\pm 2\pi/3, z) = [z - \lambda' - 3\Delta]^{-1}$. In the limit of large $J_H$ and therefore large $\Delta$, $G_{B,0} (k=\pm 2\pi/3, z) \rightarrow 0$. Therefore, the bare propagator to be used for the case of strong Hund's coupling and absent Hund's coupling has the same form: $G_{B,0} (z) = [z - \lambda]^{-1}$, where $\lambda$ has to be adjusted for the context.

\subsection{Zeroth order}

The zeroth order is readily obtained. 

\begin{align}
    \chi_{\rm sp,0} (i\nu_p) &= T\sum_n G_{B,0} (i\nu_n) G_{B,0} (i\nu_n - i\nu_p) = T\sum_n \frac{1}{i\nu_n - \lambda} \frac{1}{i\nu_n - i\nu_p - \lambda} \\  
    &= T\sum_n \frac{1}{(i\nu_n - \lambda)^2} \delta_{p,0}  = \beta n_B(\lambda) (1 + n_B(\lambda)) \delta_{p,0} \;.
\end{align}

\noindent There is thus no finite frequency impurity spin susceptibility absent hybridization with the electronic degrees of freedom. The $p=0$ (zero frequency contribution) leads to the Curie-like behavior of the static susceptibility: $\chi_{\rm sp} \simeq \beta n_B(\lambda) [ 1+ n_B(\lambda)]$.

\subsection{First order}

We then consider the first order correction to this through the addition of a self-energy bubble for the spinons, see Fig.~\ref{fig:2orders} (c). In this section, we concern ourselves with the intermediate energy window where $T_K^{\rm eff} < \omega < T_{\rm orb}$, in which the results of the preceding section are applicable. In essence, instead of considering a full $G_{\chi}$ holon propagator, we will be replacing it such that $V^2G_{\chi} = \tilde{G}_{\chi}$, i.e. a Kondo holon's Green's function. We use the instanteneous Green's function in imaginary time obtained in Eq.~\eqref{gchi-imtime}. The first order correction to the spinon bubble, as shown in Fig.~\ref{fig:2orders} (b), is 

\begin{align}
    \chi_{\rm sp,1} (\tau) &= G_{B,0} (-\tau) \int d\tau_1 d\tau_2 G_{B,0} (\tau_1) g_{c,0} (\tau_2 - \tau_1) \tilde{G}_{\chi} (\tau_2 - \tau_1) G_{B,0} (\tau - \tau_2) \\
    &= -J_K^{\rm eff} G_{B,0} (-\tau) g_{c,0} (0^+)\int d\tau_1 d \tau_2 G_{B,0} (\tau_1) G_{B,0} g_{c,0} (\tau_2 - \tau_1) \delta(\tau_2 - \tau_1) (\tau - \tau_2) \;.\\
\end{align}

\noindent In the next step, we need to use our form for a flat electronic density of states. We have that  and the resultant propagator for the conduction electrons.

\begin{equation}
    g_{c,0} (i \omega_n) =  \rho \int_{-D}^{D} d\epsilon  \frac{1}{i\omega_n-\epsilon}\;, \label{eq-gcsupp}
\end{equation}

\noindent where $\rho = \frac{1}{2D}$, and $D$ the electron bandwidth. This leads to $\rho_{c,0} (\omega) = \rho$. Because of particle-hole symmetry and the fermionic commutation relation, we then get that $g_{c,0}(\tau = 0^{+}) = -\frac12$. Therefore, after transforming to Matsubara frequencies, we obtain

\begin{align}
    \chi_{\rm sp,1} (i \nu_p) &= (J_K^{\rm eff}) T\sum_n [G_{B,0} (i\nu_n)]^2 G_{B,0} (i\nu_n - i\nu_p) \\
    &= (J_K^{\rm eff}) T\sum_n \frac{1}{(i\nu_n - \lambda)^2} \frac{1}{i\nu_n - i\nu_p - \lambda} = \frac{(J_K^{\rm eff}) n_B' (\lambda)}{i\nu_p} = \frac{\beta(J_K^{\rm eff}) n_B(\lambda) [1 + n_B (\lambda)]}{i\nu_p}\;.
\end{align}

\noindent After doing the analytic continuation $i\nu_p \rightarrow \omega + i \delta$, we see that taking the imaginary part leaves $\chi_{\rm sp, 1}''(\omega) \propto \delta(\omega)$ - there is no dissipative feature at the first order.

\subsection{Second order}

Finally, we can proceed to the second order correction to the spin susceptibility, which is outlined in Fig.~\ref{fig:2orders} (c). Again, we use the simplification of an instantaneous holon propagator after mapping the mixed valence problem onto an effective Kondo one: $\tilde{G}_{\chi}(\tau) \simeq -J_K^{\rm eff} \delta(\tau)$. The second order susceptibility becomes

\begin{align}
    \chi_{\rm sp, 2} (\tau) &= (J_K^{\rm eff})^2 G_{B,0} (-\tau) \int d\tau_1 d\tau_2 G_{B,0} (\tau_1) g_{c,0} (\tau_2 - \tau_1) g_{c,0} (\tau_1 - \tau_2) G_{B,0} (\tau_2 - \tau_1) G_{B,0} (\tau - \tau_2) \;.
\end{align}

We then Fourier transform these expressions to Matsubara frequencies. We obtain

\begin{align}
    \chi_{\rm sp, 2} (i\nu_p) &= (J_K^{\rm eff})^2 T^3 \sum_{n_1, ..., n_6} \Big[ T \int d\tau e^{(i\nu_p + i \nu_{n_1} - i \nu_{n_6}) \tau} \Big] \nonumber \\
    &\; \; \times \Big[ T \int d\tau_1 e^{(-i\nu_{n_2} + i\omega_{n_3} - i \omega_{n_4} + i \nu_{n_5}) \tau_1} \Big] \Big[ T \int d\tau_2 e^{(- i \omega_{n_3} + i \omega_{n_4} - i \nu_{n_5} + i \nu_{n_6}) \tau_2} \Big]\nonumber \\
    & \; \; \times G_{B,0} (i\nu_{n_1}) G_{B,0} (i\nu_{n_2}) g_{c,0} (i\omega_{n_3}) g_{c,0} (i\omega_{n_4}) G_{B,0} (i\nu_{n_5}) G_{B,0} (i\nu_{n_6}) \\
    &= (\rho J_K^{\rm eff})^2  \int d\epsilon d\epsilon' T^3 \sum_{n,m,q} \frac{1}{i\omega_n + i\omega_m + i\nu_p - i\nu_q - \lambda} \frac{1}{(i\omega_n + i\omega_m + i\nu_p - \lambda)^2} \frac{1}{i\nu_p - \lambda} \frac{1}{i\omega_n - \epsilon} \frac{1}{i\omega_m - \epsilon'}\;,
\end{align}

\noindent where the integral form for the conduction electron's Green's function (Eq.~\eqref{eq-gcsupp}) was included in the last line. The Matsubara sums were performed analytically on Mathematica. The result is

\begin{align}
    \chi_{\rm sp, 2} (i\nu_p) &= (\rho J_K^{\rm eff})^2  \int d\epsilon d\epsilon' \Big( \frac{\mathcal{F}_1 (\epsilon, \epsilon', \lambda, i\nu_p)}{i \nu_p}  + \frac{\mathcal{F}_2 (\epsilon, \epsilon', \lambda, i\nu_p)}{\epsilon - \epsilon' - i \nu_p}\Big)\;.
\end{align}

We are only interested in this section in a Mathematica expression for the dynamical susceptibility, i.e. $\Imp \chi_{\rm sp} (\omega)$. Therefore, we first analytically the bosonic Matsubara frequencies from the imaginary axis to the real axis: $i\nu_p \rightarrow \omega + i \delta$, and then take the imaginary part of that. We then obtain

\begin{align}
    \chi_{\rm sp, 2}'' (\omega) &= (\rho J_K^{\rm eff})^2  \int d\epsilon d\epsilon' \Big( \mathcal{F}_1 (\epsilon, \epsilon', \lambda) \delta(\omega)  + \mathcal{F}_2 (\epsilon, \epsilon', \lambda) \delta(\epsilon - \epsilon' - \omega)\Big) \;.\label{eq-imsuscSP}
\end{align}

The factors $\mathcal{F}_1$ and $\mathcal{F}_2$ are obtained from the Matsubara sums. We simply state the result here

\begin{align}
    \mathcal{F}_1 (\epsilon, \epsilon', \lambda) &= \frac{1}{\epsilon - \epsilon'} \Big( n_B(\lambda)^2 n_F'(\epsilon) - [n_F(\epsilon') - n_F(\epsilon - \lambda)] [ n_F(\epsilon) n_B'(\lambda - \epsilon + \epsilon') - n_B(\lambda - \epsilon + \epsilon') n_F'(\epsilon)] \nonumber\\ &- n_B(\lambda - \epsilon + \epsilon') n_F(\epsilon) n_F'(\epsilon') + n_B(\lambda)[n_F(\epsilon) n_F'(\epsilon)  - n_B(\lambda - \epsilon + \epsilon') n_F'(\epsilon') ]\nonumber \\
    & + n_B(\lambda)[n_F(\epsilon - \lambda)(n_B'(\lambda - \epsilon + \epsilon') + n_F'(\epsilon)) - n_F(\epsilon) (n_B'(\lambda - \epsilon + \epsilon') + n_F'(\epsilon))]\Big) \\
    \mathcal{F}_2 (\epsilon, \epsilon', \lambda) &= \frac{n_B(\lambda) - n_B(\lambda - \epsilon + \epsilon')}{(\epsilon - \epsilon')^2} [ n_B(\lambda) + n_F(\epsilon)] [n_F(\epsilon') - n_F(\epsilon - \lambda)]\;.
\end{align}

Clearly, the first term in Eq.~\eqref{eq-imsuscSP} leads to a contribution only at $\omega = 0$, and therefore does not contribute to the finite frequency behavior of the spin susceptibility. We then proceed to do the integral over $\epsilon$, which picks up the Kronecker $\delta$-function. We are therefore left with

\begin{align}
    \chi_{\rm sp, 2}'' (\omega) &= (\rho J_K^{\rm eff})^2 \frac{n_B(\lambda) - n_B(\lambda - \omega)}{\omega^2} \int  d\epsilon' [ n_B(\lambda) + n_F(\epsilon' + \omega)] [n_F(\epsilon') - n_F(\epsilon' + \omega - \lambda)] \;. \label{eq-imsuscSP2}
\end{align}

\noindent Taking the low temperature limit ($\omega/T \gg 1$), we find that $n_B(\lambda - \omega) \rightarrow -1$. Furthermore, the low-temperature substitution for the Fermi-Dirac distribution leads to 

\begin{align}
    \chi_{\rm sp, 2}'' (\omega) &= (\rho J_K^{\rm eff})^2 \frac{(1+ n_B(\lambda))}{\omega^2} \int  d\epsilon' [ n_B(\lambda) + \Theta(-\epsilon' - \omega)] [\Theta(-\epsilon') - \Theta(-\epsilon' - \omega + \lambda)] \notag \\
    &\simeq  \frac{(\rho J_K^{\rm eff} )^2 n_B(\lambda)[1+ n_B(\lambda)]}{\omega}\;, \label{eq-imsuscSP3}
\end{align}

\noindent since the integrand is only nonzero in the interval $[-\omega, 0]$, where its value is $n_B(\lambda)$. We can complete the analogy to the result of Eq. C7 in Ref.~\cite{koller2005singular} by noting that $n_B(\lambda) \simeq q = 2S/N$. 

Furthermore, connecting the formula with the one obtained in Eq.~\eqref{eqlogomega3}, we get that 

\begin{align}
    \chi_{\rm sp, 2}'' (\omega) &\simeq  \frac{n_B(\lambda)(1+ n_B(\lambda))}{\omega \Big[ \ln \Big( \frac{\omega}{T_K^{\rm eff}}\Big) \Big]^2} \;. \label{eq-imsuscSP4}
\end{align}

\noindent This completes the derivation of Eq.~(8) in the main text.

\section{V. Quasi-Power-Law Derivation}

In this section, we derive the argument given in the main text where the logarithmic corrections give rise to a quasi-power-law for a large frequency window of the intermediate regime. We start with $1/\chi''_{\rm sp} (\omega)$ and then rewrite the log, bringing back the upper frequency cutoff $\mathcal{D} = \min (T_{\rm orb}, \Gamma)$ into the expression.

\begin{align}
     1/\chi''_{\rm sp} (\omega) &= \omega \Big[ \ln \Big( \frac{\omega}{T_K^{\rm eff}} \Big) \Big]^2  = \omega \Big[ \ln \Big( \frac{\omega}{\mathcal{D}} \Big)  - \ln \Big( \frac{T_K^{\rm eff}}{\mathcal{D}} \Big) \Big]^2 \\
     &= \omega \ln^2 \Big( \frac{T_K^{\rm eff}}{\mathcal{D}}\Big) \Big[ 1 - \frac{\ln \omega/\mathcal{D}}{\ln T_K^{\rm eff}/\mathcal{D}} \Big]^2 \;. \label{eq-qpl1}
\end{align}

We compare this result with a power-law ansatz for the scaling of the susceptibility

\begin{align}
     \chi_{\rm pwl} (\omega) &= \omega^{-\gamma} \propto \frac{1}{\omega} \Big[ \Big(\frac{\omega}{\Omega} \Big)^{\epsilon/2} \Big]^{-2} = \frac{1}{\omega} \Big[ \exp \Big(\frac{\epsilon}{2} \ln \frac{\omega}{\Omega} \Big) \Big]^{-2} \;, 
\end{align}

\noindent where $\gamma = 1 + \epsilon$. We expand the exponential function, and obtain the following series

\begin{align}
     1/\chi_{\rm pwl} (\omega) &= \omega \Big[ 1 + \frac{\epsilon}{2} \ln \frac{\omega}{\Omega} +  \frac{\epsilon^2}{8} \ln^2 \frac{\omega}{\Omega} + \cdots \Big]^2 \;, \label{eq-exppw}
\end{align}

\noindent which we compare to the expression in Eq.~\eqref{eq-qpl1}. We see that, upon writing $\epsilon = -2/\ln(T_K^{\rm eff}/\mathcal{D})$, the first two terms of both expressions agree. This also brings a natural value for $\Omega = \mathcal{D}$ as a natural cutoff. The logarithmic behavior of Eq.~\eqref{eq-qpl1} will be identical to a power-law behavior with 

\begin{align}
    \gamma &= 1 + \epsilon = 1 - \frac{2}{\ln  \frac{T_K^{\rm eff}}{\mathcal{D}}} \,, \label{eq-qpwlaw}
\end{align}

\noindent if and only if $\frac{\epsilon}{2} \ln \frac{\omega}{\mathcal{D}} \ll 1$. This tells us that if the third term in Eq. \eqref{eq-exppw} is too large, then there will be noticeable differences between the logarithmic and power-law forms. For $n \simeq 2.7$, we can approximate the frequency regime where a difference would be seen. For such a parameter, we find that $\frac{\mathcal{D}}{T_K^{\rm eff}} \sim 10^4$, and therefore, that $\frac{\epsilon}{2} \ln \frac{\omega}{\mathcal{D}} \sim 1$ when $\omega \sim 10^{-3} \mathcal{D}$. This is in line with Fig.~\ref{fig:susceptibility} (c), where deviations between the two curves can be seen at low frequencies. The intermediate frequency regime, which is opened up at its widest for large Hund's coupling $J_H$ and impurities close to half-filling, therefore appears as \textit{quasi-power-law}. Only at much reduced energy scales does the logarithmic scaling of the Kondo coupling generate an upturn in the spin susceptibility.

We finish this section by illustrating the quasi-power-law exponent $\gamma$ from Eq.~\eqref{eq-qpwlaw} for the different extracted $T_K^{\rm eff}$ and $T_{\rm orb}$ as a function of $n_{\rm imp}$ from our phase diagram [Fig.~1 in the main text]. This is shown in Fig.~\ref{fig:gammanimp}.

\begin{figure}[b]
    \centering
    \includegraphics[width=0.7\linewidth]{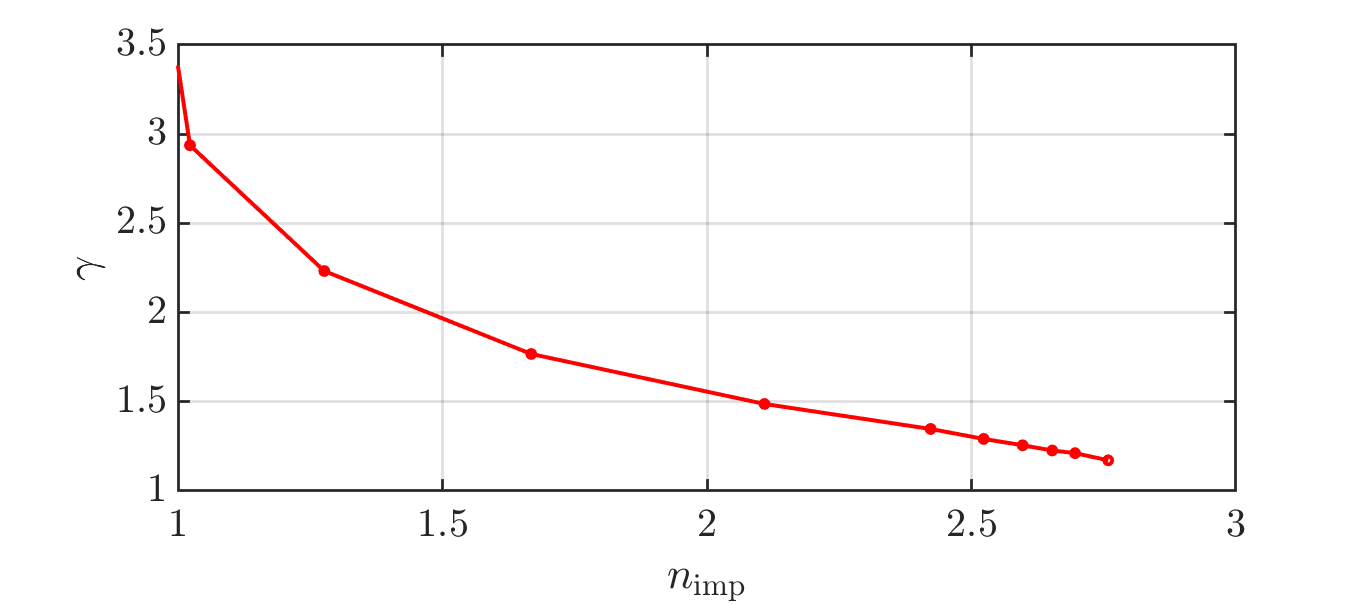}
    \caption{Value of the quasi-power-law exponent $\gamma$ from Eq.~\eqref{eq-qpwlaw}, extracted for the various impurity doping values $n_{\rm imp}$ presented in the main phase diagram and their corresponding ratios $T_K^{\rm eff}/T_{\rm orb}$. }
    \label{fig:gammanimp}
\end{figure}

As one approaches $n_{\rm imp} \sim 1.0$, the extent of the S.O.S. phase shrinks so dramatically that the assumption that $\epsilon \log \frac{\omega}{\mathcal{D}} \ll 1$ is no longer valid, and the power-law form ceases to be applicable. As $n_{\rm imp} \rightarrow 2.7$, $\gamma$ shrinks and seems to approach $1.0$ asymptotically.

\end{widetext}

\end{document}